\documentclass[11pt,prd,a4paper,preprintnumbers,amsmath,amssymb,nofootinbib]{article}
\pdfoutput=1
\usepackage{feynmp-auto,expdlist}
\usepackage{amsmath,amsfonts,amssymb}
\usepackage{graphicx}
\usepackage{enumerate}
\usepackage{hyperref}
\usepackage{latexsym}
\usepackage{hepnicenames}
\usepackage{enumerate}
\usepackage{soul}
\usepackage[normalem]{ulem}
\usepackage{wasysym}
\usepackage{makecell}
\usepackage{bbm}
\usepackage[print-unity-mantissa = false]{siunitx}

\font\tenrsfs=rsfs10 at 12pt
\font\sevenrsfs=rsfs7
\font\fiversfs=rsfs5
\newfam\rsfsfam
\textfont\rsfsfam=\tenrsfs
\scriptfont\rsfsfam=\sevenrsfs
\scriptscriptfont\rsfsfam=\fiversfs

\numberwithin{equation}{section}

\usepackage[version=4]{mhchem}
\usepackage{mathrsfs}
\usepackage{braket}
\usepackage{titling}
\usepackage{amsmath}
\usepackage{slashed}
\usepackage{amssymb}
\usepackage{epsfig}
\usepackage{graphicx}
\usepackage{color}
\usepackage{rotating}
\usepackage{hyperref}
\usepackage[margin=1.in]{geometry}
\usepackage[table,xcdraw,dvipsnames]{xcolor}
\usepackage[compress,numbers,sort]{natbib}
\usepackage{colortbl}
\usepackage{pdflscape}
\usepackage{color}
\usepackage{mathtools}
\usepackage{colortbl}
\usepackage{comment}
\usepackage{multirow}
\usepackage{booktabs}

\definecolor{nicered}{rgb}{0.7,0.1,0.1}
\definecolor{nicegreen}{rgb}{0.1,0.5,0.1}
\definecolor{red}{rgb}{1.0, 0, 0}
\definecolor{niceblue}{rgb}{0,0,0.8}
\allowdisplaybreaks

\definecolor{blus}{cmyk}{1,1,0,0.6}
\definecolor{verde}{cmyk}{0.92,0,0.59,0.25}
\definecolor{rossos}{cmyk}{0,1,1,0.55}

\hypersetup{colorlinks,bookmarksopen,bookmarksnumbered,linkcolor=blus,pdfstartview=FitH,urlcolor=rossos,citecolor=verde}

\def\eq#1{{Eq.~(\ref{#1})}}

\def\fig#1{{Fig.~\ref{#1}}}

\def\sect#1{{Section~\ref{#1}}}

\def\app#1{{Appendix~\ref{#1}}}

\renewcommand{\bar}{\overline}

\newcommand{\beq}{\begin{equation}}
\newcommand{\eeq}{\end{equation}}
\newcommand{\bea}{\begin{eqnarray}}
\newcommand{\eea}{\end{eqnarray}}

\renewcommand{\[}{\left[}

\def\be{\begin{equation}}
\def\ee{\end{equation}}

\begin{document}

\begin{center}  
{\LARGE
\bf\color{blus}
Weak nuclear decays deep-underground \\
as a probe of axion
dark matter 
} \\
\vspace{0.8cm}

{\bf Jorge Alda$^{a,b,c}$, Carlo Broggini$^{a}$, Giuseppe Di Carlo$^{d}$, Luca Di Luzio$^{a}$, \\ 
Denise Piatti$^{a,b}$, Stefano Rigolin$^{a,b}$, Claudio Toni$^{a,b,e}$ }\\[5mm]

{\it $^a$Istituto Nazionale di Fisica Nucleare, Sezione di Padova, \\
Via F. Marzolo 8, 35131 Padova (PD), Italy}\\[1mm]
{\it $^b$Dipartimento di Fisica e Astronomia `G.~Galilei', Universit\`a di Padova,
 \\ Via F. Marzolo 8, 35131 Padova (PD), Italy}\\[1mm]
{\it $^c$Centro de Astropart{\'\i}culas y F{\'\i}sica de Altas Energ{\'\i}as (CAPA), \\
Pedro Cerbuna 12,  E-50009 Zaragoza, Spain}\\[1mm]
{\it $^d$Istituto Nazionale di Fisica Nucleare, Laboratori Nazionali del Gran Sasso, \\
67100 Assergi (AQ), Italy}\\[1mm]
{\it $^e$LAPTh, Université Savoie Mont-Blanc et CNRS,
74941 Annecy, France}

\vspace{0.3cm}
\begin{quote}

We investigate the time modulation of weak nuclear decays as 
a method to probe axion dark matter. To this end, 
we develop a theoretical framework to compute the $\theta$-dependence 
of weak nuclear decays, including electron capture and $\beta$ decay, 
which enables us to predict the time variation of weak radioactivity in response to an oscillating axion dark matter background. 
As an application, we recast old data sets, from the 
weak nuclear decays of $\ce{^{40}K}$ and $\ce{^{137}Cs}$ 
taken at the underground Gran Sasso Laboratory, in order to set constraints 
on the axion decay constant, specifically 
in the axion mass range from few $10^{-23}\;$eV 
up to $10^{-19}\;$eV. 
We finally propose a new measurement at the Gran Sasso Laboratory, based on the weak nuclear decay of $\ce{^{40}K}$ via electron capture, 
in order to explore even shorter timescales, thus reaching 
sensitivities to 
axion masses up to $10^{-9}\;$eV.

\end{quote}

\thispagestyle{empty}

\end{center}

\bigskip
\setcounter{tocdepth}{2}
\tableofcontents


\section{Introduction}
\label{sec:intro}

The investigation of time-dependent nuclear decay rates traces back to the origins of radioactivity research. Notably, Maria Sk\l{}odowska-Curie’s Ph.D.~thesis~\cite{MadameCurie} described an experiment conducted to compare the radioactivity of uranium measured at midday and midnight, where no significant variation was observed. In recent years, some studies have reported small periodic modulations in the decay rates of various nuclei at the per mille level, with timescales ranging from days to months and even up to a year (see~\cite{McDuffie:2020uuv} and references therein). However, other researchers have challenged these findings, providing no evidence for such effects~\cite{Pomme1,Pomme2,Pomme3}.

To shed light on this debated phenomenon, 
a set of precise $\gamma$-spectroscopy experiments 
were conducted
at the underground Gran Sasso Laboratory~\cite{Bellotti:2012if,Bellotti:2015toa,Bellotti:2013bka,Bellotti:2018jzd}. The unique advantages of this facility, such as the rock overburden that significantly reduces 
the cosmic muon and neutron fluxes,  
allowed the authors of Refs.~\cite{Bellotti:2012if,Bellotti:2015toa,Bellotti:2013bka,Bellotti:2018jzd} to minimize external influences. This suppression effectively eliminates the contribution of cosmic ray flux variations, which typically exhibit annual modulations with amplitudes of a few percent~\cite{Bellotti:2015toa}. Those measurements ultimately ruled out modulations of the decay constants in several radioisotopes, such as $\ce{^{137}Cs}$~\cite{Bellotti:2012if}, $\ce{^{222}Rn}$~\cite{Bellotti:2015toa}, $\ce{^{232}Th}$~\cite{Bellotti:2013bka}, $\ce{^{40}K}$ 
and $\ce{^{226}Ra}$~\cite{Bellotti:2018jzd}, with amplitudes exceeding a few parts per $10^5$ for periods between a few hours and a year.

The aforementioned limits on the time modulation of radioisotope decay rates can also be employed to derive stringent constraints on the time variation of fundamental constants of nature, potentially arising from underlying fundamental physics. A particularly motivated 
framework is that of the Quantum Chromodynamics (QCD) axion~\cite{Peccei:1977hh,Peccei:1977ur,Weinberg:1977ma,Wilczek:1977pj}, which simultaneously addresses both the strong CP problem 
and the dark matter puzzle~\cite{Dine:1982ah,Abbott:1982af,
Preskill:1982cy}. For a review of recent experimental strategies for axion detection, see e.g.~\cite{Irastorza:2018dyq,DiLuzio:2020wdo,Sikivie:2020zpn}.
In particular, promoting the QCD $\theta$-term 
to a time-varying 
axion 
field 
enables a model-independent test of the axion-gluon coupling through the oscillating electric dipole moment (EDM) of the neutron,  
induced by the axion dark matter background~\cite{Graham:2013gfa,Budker:2013hfa,Stadnik:2013raa}. 
Other approaches to set laboratory limits on the axion-gluon coupling have been discussed 
\emph{e.g.}~in Refs.~\cite{Abel:2017rtm,Roussy:2020ily,Schulthess:2022pbp,JEDI:2022hxa,Zhang:2022ewz,Madge:2024aot,Fan:2024pxs}. 

Recently, the authors of Ref.~\cite{Zhang:2023lem} have proposed to look for the time variation  
of the decay rate of certain radioisotopes, focussing 
on the $\theta$-dependence 
of $\beta$ decay,  
previously developed in~\cite{Lee:2020tmi}.  
This allowed them to set bounds on the axion coupling to gluons from tritium decay, 
based on data taken at the European Commission's Joint Research Centre~\cite{Pomme2}. 

Following a similar approach, 
Ref.~\cite{Broggini:2024udi} developed a theoretical 
framework 
to describe the $\theta$-dependence 
of $\alpha$-decays, enabling predictions for the time variation of $\alpha$-radioactivity induced by an oscillating axion dark matter background. Focussing on the $\alpha$-decay of $\ce{^{241}Am}$, 
the authors of Ref.~\cite{Broggini:2024udi} proposed and deployed 
a prototype experiment, RadioAxion-$\alpha$, deep underground at the Gran Sasso Laboratory, collecting data since the summer of 2024.

In this work, we extend the aforementioned program by investigating the $\theta$-dependence of weak nuclear decays, focusing specifically on electron capture (EC) and $\beta$ decay. This analysis enables us to constrain the axion parameter space by reinterpreting the datasets from Ref.~\cite{Bellotti:2018jzd} ($\ce{^{40}K}$ decay via EC, covering periods between 6 hours and 800 days) and Ref.~\cite{Bellotti:2012if} ($\ce{^{137}Cs}$ $\beta$ decay, covering periods between 6 hours and 1 year). Additionally, we propose a new measurement at the Gran Sasso Laboratory, based on the weak nuclear decay of $\ce{^{40}K}$ via EC, 
in order to explore timescales down to \qty{1}{\micro\second}.

This work is structured as follows. In \sect{sec:theta_pion} we introduce the formalism needed to estimate the 
$\theta$-dependence of nuclear quantities. 
Next, in \sect{sec:signal} we provide a general overview of weak 
nuclear decays, which allows us to derive in \sect{sec:timemodwnd}
the time modulation for EC and $\beta$ decay, induced by axion dark matter. 
In \sect{sec:expanddata} we describe the experimental setups of Refs.~\cite{Bellotti:2018jzd,Bellotti:2012if} 
and discuss the inferred limits 
on the axion dark matter parameter space. We then describe the future developments of the  
$\ce{^{40}K}$ experiment and we finally 
conclude in \sect{sec:concl}, while more technical details on the 
calculation of weak nuclear decays are deferred to \app{AppA}.

\section{\texorpdfstring{\boldmath{$\theta$}}{θ}-dependence of mesons, nucleons and nuclei}
\label{sec:theta_pion}

The QCD $\theta$-term is defined by the operator  
\beq 
\label{eq:deftheta}
\mathcal{L}_\theta = 
\frac{g_s^2}{32\pi^2} \theta
G^a_{\mu\nu} \tilde G^{a\,\mu\nu} \, , 
\eeq
with the present constraint, $|\theta|\lesssim 10^{-10}$, 
arising from the non-observation of the 
neutron EDM~\cite{Abel:2020pzs}. 
The smallness of $\theta$ constitutes the so-called strong CP problem, 
which can be solved by promoting the 
$\theta$-term to be a dynamical field, $\theta \to a(x)/f_a$, 
where $a(x)$ is the axion 
field and $f_a$ the axion decay constant. 
The axion field acquires a potential in the background 
of QCD instantons and relaxes dynamically to the 
CP-conserving minimum, 
thus explaining the absence of CP violation in strong interactions~\cite{Peccei:1977hh,Peccei:1977ur,Weinberg:1977ma,Wilczek:1977pj}. 

In the following, we will be interested in the $\theta$-dependence of nuclear quantities, 
anticipating the fact that we will interpret $\theta(t)$ as a time-varying background axion field, 
related to 
the dark matter of the universe~\cite{Dine:1982ah,Abbott:1982af,
Preskill:1982cy}.
The consequences of a non-zero $\theta$ in nuclear physics have been previously investigated in 
Refs.~\cite{Ubaldi:2008nf,Lee:2020tmi}, also in connection with the idea of establishing an anthropic bound on $\theta$. 

There are various ways in which the $\theta$-dependence can manifest in nuclear physics, the most prominent is 
through the pion mass~\cite{Leutwyler:1992yt,Brower:2003yx}
\beq 
M^2_{\pi}(\theta) = M^2_{\pi} \cos{\frac{\theta}{2}} \sqrt{1 + \varepsilon^2 \tan^2\frac{\theta}{2}} \, , 
\eeq
with $M_{\pi} = \qty{139.57}{\mega\electronvolt}$ and $\varepsilon = (m_d-m_u)/(m_d+m_u)$.
The $\theta$-dependence of other 
low-lying resonances, including $\sigma(550)$, $\rho(770)$ and $\omega(782)$ --  
which, along with the pion, are  
responsible for the mediation of nuclear forces in the one-boson-exchange 
approximation --  
has been determined based on $\pi\pi$ scattering data
in Ref.~\cite{Acharya:2015pya}.
The $\theta$-dependence of the neutron-proton mass difference has been instead determined from the next-to-leading order chiral pion-nucleon Lagrangian and gives at lowest order in the pion mass~\cite{Lee:2020tmi}
\be
\label{eq:Deltampn}
\Delta m_N (\theta) \equiv m_n(\theta) - m_p(\theta) =-4c_5 \varepsilon \frac{M_\pi^4}{M_\pi^2(\theta)}
 \ ,
\ee
where 
$c_5 = \qty[separate-uncertainty=true]{-0.074\pm0.006}{\giga\electronvolt^{-1}}$
is a low-energy constant of the chiral Lagrangian.

A key role 
for the binding energy
of heavy nuclei 
is played by the $\sigma$ and $\omega$ channels, 
via the contact interactions~\cite{Furnstahl:1999rm}
\beq 
H_{\rm strong} \supset G_S (\bar N N) (\bar N N) + G_V (\bar N \gamma_\mu N) (\bar N \gamma^\mu N) \, , 
\eeq
which control, respectively, the scalar (attractive) and vector (repulsive) part of the nucleon-nucleon interaction~\cite{Donoghue:2006rg,Damour:2007uv}. 
To describe their $\theta$-dependence we employ the following parametrization
\beq 
\eta_S = \frac{G_S(\theta)}{G_S(\theta=0)} \, , \qquad 
\eta_V = \frac{G_V(\theta)}{G_V(\theta=0)} \, .
\eeq
In Ref.~\cite{Donoghue:2006du} it was found that 
the pion mass dependence of $\omega$ exchange 
leads to subleading corrections compared to the effects related to the 
$M^2_{\pi}$ sensitivity of the scalar channel. Hence, to a good approximation, we can take $\eta_V = 1$ 
and consider only the leading $\theta$-dependence in the scalar channel, 
which is described by the following fit~\cite{Ubaldi:2008nf} to 
Fig.~2 in~\cite{Damour:2007uv} 
\beq
\label{eq:etaStheta}
\eta_S (\theta) = 1.4 - 0.4 \frac{M^2_{\pi}(\theta)}{M^2_{\pi}} \, .
\eeq
The binding energy (BE) for a heavy nucleus of mass number $A$ and atomic number $Z$ can be described as a sum of contributions, each accounting for a different physical effect, given by the semi-empirical mass formula~\cite{Weizsacker:1935bkz,book:80102} 
\be
\label{eq:semi_emp_mass_formula}
\text{BE}=
a_V A - a_S A^{2/3} -a_C\frac{Z(Z-1)}{A^{1/3}}-a_A\frac{(A-2Z)^2}{A}+\frac{a_P}{A^{1/2}}\delta(A,Z) \ ,
\ee
where $\delta(A,Z)=+1(-1)$ 
for even $A$ with an even (odd) value of $Z$, 
while it vanishes for odd $A$. 
The values of the (positive-defined) coefficients are obtained at $\theta=0$ from a fit on the experimental data, see \emph{e.g.} Ref.~\cite{GJORGIEVSKA2024113403}.
These terms represent, respectively:
\begin{itemize}
\item a volume contribution, hence proportional to the mass number, given by $a_V A$;
\item a negative surface contribution, roughly equivalent to liquid surface tension, given by $-a_S A^{2/3}$;
\item a repelling Coulomb energy $-a_C Z(Z-1)/A^{1/3}$;
\item an asymmetry term which accounts for the Pauli exclusion principle, since unequal numbers of neutrons and protons imply filling higher energy levels for one type of particle, that goes as $-a_A (A-2Z)^2/A$;
\item a pairing term, accounting for the tendency of proton pairs and neutron pairs to occur, which scales as $A^{-1/2}$ and hence is typically negligible for large $A$ nuclei.
\end{itemize}
Based on the relativistic mean-field simulations of~\cite{Furnstahl:1999rm} for 
two specific nuclei ($\ce{^{16}O}$ and $\ce{^{208}Pb}$), Ref.~\cite{Damour:2007uv} 
extrapolated the variation of the BE
for a general heavy nucleus as (keeping only the variation due to $\eta_S(\theta)$) 
\be
\label{eq:BEtheta}
\delta\text{BE}(\theta)\equiv\text{BE}(\theta)-\text{BE}(\theta=0) =(120 A - 97 A^{2/3})(\eta_S (\theta) -1) \, \text{MeV} \, , 
\ee
where the terms proportional to $A$ and $A^{2/3}$ represent 
the volume and surface terms, 
in analogy to the semi-empirical mass formula in~\eq{eq:semi_emp_mass_formula}, 
which typically capture the leading contribution to the BEs. 

Note, however, that the energy released in weak nuclear decays depends on the difference of BEs of nuclei with the same mass number $A$, thus implying that the variation due to Eq.~\eqref{eq:BEtheta} disappears in the difference between the initial and final nuclear states. 
Subleading contributions from the remaining terms 
in~\eq{eq:BEtheta}, 
which also depend on the atomic number $Z$, might then become relevant, but no estimate of their $\theta$-dependence has been performed so far, at least for large-$A$ nuclei which are the main focus of the present work.\footnote{In the context of the $\beta$ decay of low-$A$ nuclei like tritium, Ref.~\cite{Zhang:2023lem} introduced the $\theta$-dependence of the BEs  
relying on 
the explicit calculation 
of Ref.~\cite{Lee:2020tmi} for $A \leq 4$ systems, 
also based on the numerical results of Ref.~\cite{Gattobigio:2012tk}.} Hence, 
we conservatively neglect the latter contributions, 
and include the proton-neutron mass difference in~\eq{eq:Deltampn} 
as the main source of $\theta$-dependence in weak nuclear decays.

\section{Overview of weak nuclear decays}
\label{sec:signal}

As a preliminary ingredient to describe the $\theta$-dependence 
of weak nuclear decays, we introduce in this section the multipole expansion formalism within the context of semi-leptonic weak processes like electron capture and $\beta$ decay, 
similarly to the approach taken in Ref.~\cite{Barducci:2022lqd}.

\subsection{Nuclear states and weak interactions}

We describe the weak interaction of a semi-leptonic process through the Hamiltonian
\be
H_{\rm weak}=-\frac{G_F}{\sqrt{2}}\int\!d^{3}\vec{r}\,\mathcal{J}_{\mu}\!(\vec{r}) j^{\mu}\!(\vec{r}) \, ,
\ee
where $G_F=\qty{1.1668e-5}{\giga\electronvolt^{-2}}$ is the Fermi constant, the nuclear current $\mathcal{J}^{\mu}=(\mathcal{J}^{0},\vec{\mathcal{J}})$ encodes quantum operators containing all the information of the nuclear matter fields while $j^{\mu}$ is the lepton weak current.
At the lowest order in the interaction picture, the nuclear matrix element is given by
\be
\label{matrix-element-weak}
\mathcal{T}_{fi}=
\braket{f|H_{\rm weak}|i}=
-\frac{G_F}{\sqrt{2}}\bra{N_f}\int\!d^{3}\vec{r} \, \,\mathcal{J}_{\mu}\!(\vec{r}) e^{-i\vec{k}\cdot\vec{r}} \, l^{\mu} \ket{N_i} \, ,
\ee
where $\vec k$ is the exchanged momentum and 
$l^\mu=(l_t,\vec{l})$ is the the fermionic current in momentum space, defined through
\be
\braket{f| j^\mu (\vec{r}) |i}_{\rm lept.} = l^\mu e^{-i\vec{k}\cdot\vec{r}} \ ,
\ee
while $\ket{N_i}=\ket{J_{i}M_{i}}$ and $\ket{N_f}=\ket{J_{f}M_{f}}$ 
indicate the nuclear matter initial and final states, 
as labeled by their total angular momentum $J_{i(f)}$ 
and its projection $M_{i(f)}$. 
In the following, we want to compute the decay widths of semi-leptonic processes in order to evaluate the impact of a non-vanishing $\theta$-term. To perform the calculation, it will turn out to be useful to expand the nuclear matrix elements in terms of spherical tensor operators through a multipole expansion.

\subsection{Multipole expansion}

Spherical operators ${\cal O}_{JM}$ are irreducible tensor operators which satisfy the Wigner-Eckart theorem~\cite{book:17167}
\be
\begin{split}\label{eq:wigner}
\braket{J_{f}M_{f}|{\cal O}_{J,-M}|J_{i}M_{i}}=\frac{(-1)^{J_{i}-M_{i}}}{\sqrt{2J+1}}
C^{J,-M}_{J_f ,M_f ;J_i, -M_i}
\braket{J_{f}||{\cal O}_{J}||J_{i}}.
\end{split}
\ee
The reduced matrix element $\braket{J_{f}||{\cal O}_{J}||J_{i}}$ contains all the physical information of the operator while its behavior under rotation is completely set by the Clebsh-Gordan coefficient  $C^{J,-M}_{J_f ,M_f ;J_i, -M_i}$. We define the spherical 
operators\footnote{$Y_{JM}$, $j_{J}(x)$ and $\textbf{Y}_{J\ell M}(\hat{r})$ are respectively the 
spherical harmonics, the vector Bessel functions and the vector spherical harmonics, see \emph{e.g.}~\cite{book:17167}. 
Explicit expressions in the long wavelength approximation are provided in \app{AppA}.} 
\begin{align}
\mathcal{M}_{JM} & =\int d^{3}\vec{r} \ j_{J}(kr)Y_{JM}(\hat{r})\mathcal{J}^{0}(\vec{r}) \ , \label{mso1}\\
\mathcal{L}_{JM}  &=\frac{i}{k}\int d^{3}\vec{r} \ \vec{\nabla}[j_{J}(kr)Y_{JM}(\hat{r})]\cdot\vec{\mathcal{J}}(\vec{r})\ , \label{mso2}\\
\mathcal{T}_{JM}^{\rm el} & =\frac{1}{k}\int d^{3}\vec{r} \ \vec{\nabla}\times[j_{J}(kr)\textbf{Y}_{JJM}(\hat{r})]\cdot\vec{\mathcal{J}}(\vec{r}) \ ,\label{mso3}\\
\mathcal{T}_{JM}^{\rm mag} & =\int d^{3}\vec{r} \ [j_{J}(kr)\textbf{Y}_{JJM}(\hat{r})]\cdot\vec{\mathcal{J}}(\vec{r}) \ , \label{mso4}
\end{align}
where $r = |\vec r|$, $k = |\vec k|$ and expand the nuclear matrix elements as a sum of reduced matrix elements of spherical operators.
We also introduce the notation
\be
l_z \equiv \hat{e}_z \cdot \vec{l} \qquad \text{and} \qquad l_{\pm}\equiv \mp\frac{\hat{e}_x \pm i \hat{e}_y}{\sqrt{2}} \cdot \vec{l} \, ,
\ee
where $\hat{e}_{x,y,z}$ is an orthonormal tridimensional basis such that $\hat{e}_{z}=\vec{k}/k$.
In the case of interest, one finds
\begin{align}
\mathcal{T}_{fi}=&\sum_{\substack{J\geq 0,\\ |M|\leq J}}(-i)^{J}\sqrt{4\pi} C^{J,-M}_{J_f, M_f; J_{i}, -M_{i}} \braket{N_f||\left[ l_t\mathcal{M}_{J}-l_z \mathcal{L}_{J}\right]||N_i}D_{-M,0}^{(J)}(\phi,\theta,\beta) \nonumber \\ 
-& \sum_{\substack{J\geq 1,\\ |M|\leq J,\\ \lambda=\pm1}}(-i)^{J}\sqrt{2\pi}l_{\lambda} C^{J,-M}_{J_f, M_f; J_{i}, -M_{i}} \braket{N_f||\left[ \mathcal{T}_{J}^\text{el}+\lambda\mathcal{T}_{J}^\text{mag}\right]||N_i} D_{-M,-\lambda}^{(J)}(\phi,\theta,\beta) \ ,
\end{align}
where the indices $J$ and $M$ denote the total angular momentum of the weak lepton current and its projection.
The rotation $D$ matrices (defined as in Ref.~\cite{book:17167}) play the role of the wave function, whose modulus squared gives the probability to have the exchanged momentum $\vec{k}$ along the $(\phi,\theta)$ direction, with $\beta$ defining a rotation along this direction.

By explicit calculation, the squared amplitude with a sum over the nuclear polarizations gives
\begin{align}
\frac{1}{2J_i+1}\sum_{M_f M_i} & \left| \braket{f|H_{\rm weak} |i} \right|^2  = \frac{G^2_F}{2}\frac{4\pi}{2J_{i}+1} \nonumber \\
\times\Biggl\{ &\sum_{J\geq 0}
\left[
l_t l_t^* \left|\braket{N_f||\mathcal{M}_{J}||N_{i}}\right|^{2} + l_z l_z^* \left|\braket{N_f||\mathcal{L}_{J}||N_{i}}\right|^{2} \right] \nonumber \\
-&\sum_{J\geq 0}
\left[
2\,\text{Re}\!\left(l_t l_z^*\braket{N_f||\mathcal{M}_{J}||N_{i}}\braket{N_f||\mathcal{L}_{J}||N_{i}}^*\right)
\right]
\nonumber \\
+ & \sum_{J\geq 1}\left[\frac{1}{2}\left(\vec{l}\cdot\vec{l}^* - l_z l_z^*\right)\left(\left|\braket{N_f||\mathcal{T}_{J}^\text{el}||N_{i}}\right|^{2} + \left|\braket{N_f||\mathcal{T}_{J}^\text{mag}||N_i}\right|^{2}\right) \right] \nonumber \\
-&\sum_{J\geq 1}\left[ i (\vec{l}\times\vec{l}^*\cdot\hat{e}_z) \, \text{Re}\!\left(\braket{N_f|| \mathcal{T}_{J}^\text{el}||N_{i}}\braket{N_f||\mathcal{T}_{J}^\text{mag}||N_{i}}^* \right) \right] \Biggr\} \ . \label{dl1}
\end{align}
This is a general result and holds for any semi-leptonic nuclear processes. In addition to lowest-order perturbation theory in the weak coupling constant $G_F$, it assumes only the existence of a local weak nuclear current operator and that the initial and final nuclear states are eigenstates of angular momentum~\cite{book:80102}.

\subsection{Selection rules}
\label{sec:selrul}

The angular momentum conservation law, encoded in the Clebsh-Gordan coefficient of Eq.~\eqref{eq:wigner}, states that the matrix element of the spherical operators vanishes unless the following conditions are satisfied
\begin{align}
& |J_{f}-J_{i}|\leq J\leq J_{f}+J_{i}\ , \nonumber \\
& M=M_{i}-M_{f} \ .
\end{align}
Moreover, an additional selection rule comes from the requirement of parity conservation. Under parity the spherical operators we introduced transform as
\begin{align}
\mathcal{M}_{JM} & \to \pi_{\mathcal{J}}(-1)^{J+1} \mathcal{M}_{JM} \ , \label{mso1p}\\
\mathcal{L}_{JM}  & \to \pi_{\mathcal{J}}(-1)^{J+1} \mathcal{L}_{JM} \ , \label{mso2p}\\
\mathcal{T}_{JM}^{\rm el} &  \to \pi_{\mathcal{J}}(-1)^{J+1} \mathcal{T}_{JM}^{\rm el} \ ,\label{mso3p}\\
\mathcal{T}_{JM}^{\rm mag} &  \to \pi_{\mathcal{J}}(-1)^{J}  \mathcal{T}_{JM}^{\rm mag} \ , \label{mso4p}
\end{align}
where $\pi_{\mathcal{J}}$ is the nuclear current parity. Then parity conservation enforces $\braket{N_f|\mathcal{O}_{JM}|N_i}=0$ unless $\Delta\pi\equiv \pi_i \pi_f=\pi_\mathcal{O}$ for any of the operators.

\subsection{Semi-leptonic processes}

While Eq.~\eqref{dl1} holds for a generic semi-leptonic processes, the actual expression of the lepton current $l^\mu$ and the total decay width depend on the particular process. Here we are interested in the electron capture and in the $\beta$ decay 
processes.
 
\subsubsection{Electron capture}

This process takes place when a nucleus absorbs an electron lying on the lowest atomic orbital and decays emitting a neutrino, explicitly
\be
e^- + N_i(A,Z) \to N_f(A,Z-1) + \nu \ .
\ee
The weak lepton current is then given by
\be
l^\mu = \bar{u}(p_\nu)\gamma^\mu (1-\gamma_5)u(p_e) \frac{\psi_{1s}(\vec{r})}{\psi_{\vec{p}_e}(\vec{r})} \ ,
\ee
where we divided by the free wave function $\psi_{\vec{p}_e}(\vec{r})=e^{i\vec{p}_e \cdot\vec{r}}$ and multiplied by the orbital wave function $\psi_{1s}(\vec{r})$ to take into account the fact that the initial electron is in a bound state. However, the nucleus has actually a finite extent and its radius is typically much smaller than the Bohr radius. This means that only the wave function near the origin contributes to the matrix element in Eq.~\eqref{matrix-element-weak}, \emph{i.e.} we can approximate
\be
\psi_{1s}(\vec{r})\approx\psi_{1s}(0)= \frac{1}{\sqrt{\pi}} (Z\alpha_{\rm QED} m_e)^{3/2} \ .
\ee
The energy conservation condition reads
\be
\label{eq:enconsEC}
m_e - E_{1s} + m_{i} = m_{f} + E_\nu \ ,
\ee
where the orbital binding energy $E_{1s}\sim \mathcal{O}(Z^2 \times \qty{10}{\electronvolt})$ is typically negligible, as well as the electron momentum.
Thus the unpolarized total decay width for the electron capture process is given by~\cite{book:80102}
\begin{align}
\label{eq:GammaEC}
\Gamma_{\rm EC} = \frac{G^2_F E^2_\nu}{2\pi}\frac{4\pi}{2J_i+1}\Biggl\{ &\sum_{J\geq 0}  \left|\braket{N_f||\mathcal{M}_{J}-\mathcal{L}_{J}||N_{i}}\right|^{2} \nonumber \\
+&\sum_{J\geq 1} \left|\braket{N_f||\mathcal{T}_{J}^{el}-\mathcal{T}_{J}^{mag}||N_{i}}\right|^{2} \Biggr\} |\psi_{1s}(0)|^2 \ . 
\end{align}

\subsubsection{$\beta$ decay}

The $\beta$ decay nuclear process is given by
\be
N_i(A,Z) \to \begin{cases}
N_f(A,Z-1)+e^+ +\nu \quad \text{for $\beta^+$ decay \, ,} \\
N_f(A,Z+1)+e^- +\bar\nu \quad \text{for $\beta^-$ decay \, ,}
\end{cases}
\ee
with the following weak lepton current
\be
l^\mu=\begin{cases}
\bar{u}(p_\nu)\gamma^\mu(1-\gamma_5)v(p_{e}) \quad \text{for $\beta^+$ decay \, ,} \\
\bar{u}(p_e)\gamma^\mu(1-\gamma_5)v(p_{\nu}) \quad \text{for $\beta^-$ decay \, .}
\end{cases}
\ee
The energy conservation condition reads
\be
m_i=m_f+E_e+E_{\nu} \ ,
\ee
and we introduce the notation
\be
v_e\equiv\frac{|\vec{p}_e|}{E_e}=\sqrt{1-\frac{m_e^2}{E_e^2}} \, ,
\quad
\cos\theta_e\equiv\frac{\vec{k}\cdot\vec{p}_e}{|\vec{k}| |\vec{p}_e|} \, ,
\quad
\cos\theta_\nu\equiv\frac{\vec{k}\cdot\vec{p}_\nu}{|\vec{k}| |\vec{p}_\nu|} \, , 
\ee
from which
\begin{gather}
\cos\theta_{e\nu}\equiv\frac{\vec{p}_\nu\cdot\vec{p}_e}{|\vec{p}_\nu| |\vec{p}_e|}=\cos\theta_e \cos\theta_\nu + \sin\theta_e \sin\theta_\nu \cos(\phi_e - \phi_\nu) \ , \\
k=|\vec{p}_e+\vec{p}_\nu|=\sqrt{E_\nu^2+E_e^2-m_e^2+2E_\nu\sqrt{E_e^2-m_e^2}\cos\theta_{e\nu}} \ ,
\end{gather}
where $\phi_{e}$ and $\phi_\nu$ are defined as the azimuthal angles of the charged and neutral lepton momentum in a system where $\vec{k}$ is along the $z$ axis.
Thus the unpolarized differential decay width is found to be~\cite{book:80102}
\begin{align}
\label{eq:Gamma:beta}
d\Gamma_{\beta^\pm}=\frac{G^2_F}{2\pi^3} v_e & E^2_e (m_i - m_f -E_e)^2  dE_e \frac{d\Omega_e}{4\pi} \frac{d\Omega_{\nu}}{4\pi} \frac{4\pi}{2J_{i}+1} \nonumber \\
\times\Biggl\{ &\sum_{J\geq 0}
\left[
(1+v_e\cos\theta_{e\nu}) \left|\braket{N_f||\mathcal{M}_{J}||N_{i}}\right|^{2} \right] \nonumber \\
+ &\sum_{J\geq 0} \left[ (1-v_e\cos\theta_{e\nu} +2v_e\cos\theta_{e}\cos\theta_{\nu}) \left|\braket{N_f||\mathcal{L}_{J}||N_{i}}\right|^{2} \right] \nonumber \\
-&\sum_{J\geq 0}
\left[
2(\cos\theta_\nu+v_e\cos\theta_e)\,\text{Re}\!\left(\braket{N_f||\mathcal{M}_{J}||N_{i}}\braket{N_f||\mathcal{L}_{J}||N_{i}}^*\right)
\right]
\nonumber \\
+ & \sum_{J\geq 1}\left[(1-v_e\cos\theta_e\cos\theta_\nu)\left(\left|\braket{N_f|| \mathcal{T}_{J}^\text{el}||N_{i}}\right|^{2} + \left|\braket{N_f||\mathcal{T}_{J}^\text{mag}||N_i}\right|^{2}\right) \right] \nonumber \\
\mp&\sum_{J\geq 1}\left[(\cos\theta_\nu - v_e\cos\theta_e) \, \text{Re}\!\left(\braket{N_f|| \mathcal{T}_{J}^\text{el}||N_{i}}\braket{N_f||\mathcal{T}_{J}^\text{mag}||N_{i}}^* \right) \right] \Biggr\} F(Z,E_e) \ ,
\end{align}
where $d\Omega_{e(\nu)}=d\phi_{e(\nu)} d\cos\theta_{e(\nu)}$ while $F(Z,E_e)$ is a form factor that takes into account that in the final state the charged lepton and the nucleus are electromagnetically interacting. An approximate treatment of the Coulomb interaction is obtained by multiplying the decay width by the ratio of the Coulomb and free squared wave function at $\vec{r}\approx0$, since the nuclear radius is much smaller than the lepton wavelength, which is given by~\cite{book:80102}
\be
F(Z,E_e)\equiv\left|\frac{\psi_{\vec{p}_e}(0)_{\rm Coul.}}{\psi_{\vec{p}_e}(0)}\right|^2=\frac{2\pi\xi}{e^{2\pi\xi}-1} \ ,
\ee
with $\xi=\pm Z\alpha_\text{QED}/v_e$ for the $\beta^\pm$-decay.

\section{Time modulation of weak nuclear decays}
\label{sec:timemodwnd}

With the results of Sec.~\ref{sec:theta_pion} and the formalism developed in Sec.~\ref{sec:signal}, we will now proceed to derive 
the theoretical sensitivity of the weak nuclear decays of interest to a time-varying $\theta$-term.

Assuming an oscillating axion dark matter field from misalignment~\cite{Dine:1982ah,Abbott:1982af,
Preskill:1982cy}, 
the time dependence of the $\theta$ angle can be approximated as
$\theta(t) \approx \theta_0 \cos(m_a t)$, 
where we neglected non-relativistic corrections 
to the axion dispersion relation,  
and
\begin{align}
\label{eq:theta0DM}
\theta_0&=\frac{\sqrt{2\rho_{\rm DM}}}{m_a f_a} \, , 
\end{align}
in terms of $\rho_{\rm DM} \approx \qty{0.45}{\giga\electronvolt/\centi\meter^3}$.
For
a standard QCD axion, one has 
\beq 
\label{eq:QCDband}
m_a f_a = \frac{\sqrt{m_u m_d}}{m_u + m_d} m_\pi f_\pi = (\qty{76}{\mega\electronvolt})^2 \, , 
\eeq
corresponding to $\theta_0 = \num{5.5e-19}$. In the following, we will treat $m_a$ and $f_a$ as independent parameters and discuss the sensitivity 
of weak decay observables in the $(m_a, 1/f_a)$ plane. 

Following Refs.~\cite{Zhang:2023lem,Broggini:2024udi}, we introduce the 
observable
\beq 
\label{eq:Itdef}
I(t) \equiv \frac{\Gamma_{\rm weak}(\theta(t)) - \langle \Gamma_{\rm weak} \rangle }{\langle \Gamma_{\rm weak} \rangle } \, , 
\eeq
where $\langle \Gamma_{\rm weak} \rangle$ denotes a time average.  
Given that the main 
$\theta$-dependence in~\eq{eq:Deltampn} and (\ref{eq:etaStheta}) arises through the pion mass, 
we expect that 
$\Gamma_{\rm weak}(\theta)$ is analytic in $\theta^2$ and 
admits the Taylor expansion
\beq 
\label{eq:Texpansion}
\Gamma_{\rm weak}(\theta) \approx \Gamma_{\rm weak}(0) + \mathring{\Gamma}_{\rm weak}(0) \theta^2 \, , 
\eeq 
where we introduced the derivative symbol, $\mathring{f} \equiv df/d\theta^2$. 
Since $\theta^2 \ll 1$,~\eq{eq:Texpansion} does provide an excellent approximation to the full 
$\theta$-dependence. 
Using $\langle \cos^2 (m_a t) \rangle = 1/2$ 
and expanding at the first non-trivial order in $\theta_0$, 
we find 
\begin{align} 
\label{eq:Itpred}
I(t) \approx& \frac{1}{2} \frac{\mathring{\Gamma}_{\rm weak}(0)}{\Gamma_{\rm weak}(0)} \theta_0^2 \cos(2 m_a t) \nonumber \\
=& 3.4 \times 10^{-8} \cos(2 m_a t) \left( \frac{\rho_{\rm DM}}{\qty{0.45}{\giga\electronvolt/\centi\meter^3} } \right) 
\left( \frac{\qty{1e-16}{\electronvolt}}{m_a} \right)^2  \left( \frac{\qty{1e8}{\giga\electronvolt}}{f_a} \right)^2 \left(\frac{\mathring{\Gamma}_{\rm weak}(0)}{\Gamma_{\rm weak}(0)}\right)
\,.
\end{align}
Note that the main $\theta$-dependence arises from:
\begin{itemize}
\item the masses of the nuclei, \emph{i.e.} the kinetic energy released to the leptons in the final state;
\item the reduced nuclear matrix elements $\braket{N_f||\mathcal{O}||N_i}$.
\end{itemize}
The calculation of the latter is a difficult task 
that lies beyond the scope of this work.
However, the $\theta$-dependence of the released kinetic energy 
captures the leading-order effects we are investigating, particularly in scenarios where the effect is small, making the decay width highly sensitive to even minor variations in the phase space. Therefore, in the following analysis, 
we will disregard the $\theta$-dependence of the reduced nuclear matrix elements.

In the subsequent subsections, 
we will specialize the general formulae derived above to two cases of interest: the 
$\ce{^{40}K}$ decay via EC and the $\ce{^{137}Cs}$ $\beta$ decay.  
These cases will enable us to reinterpret existing datasets, 
respectively from Ref.~\cite{Bellotti:2018jzd} and~\cite{Bellotti:2012if}, in order to set constraints on the axion dark matter parameter space.

\subsection[$\ce{^{40}K}$ electron capture decay]{\texorpdfstring{\boldmath{$\ce{^{40}K}$}}{40K} electron capture decay}

The nucleus of
$\ce{^{40}K}$
decays to the first excited state of argon
$\ce{^{40}Ar^\star}$
through electron capture the $10.31(4)\%$ of the time~\cite{Wang:2021xhn,KDK:2022hgg}.
The latter then decays to the ground state while emitting a photon with energy $\qty{1460.820(5)}{\kilo\electronvolt}$.
The quantum numbers of the initial and final nuclear state are respectively $J_i^{\pi_i}=4^-$ and $J_f^{\pi_f}=2^+$, thus the selection rules discussed in \sect{sec:selrul}
enforce the restriction $2\leq J\leq6$ on the multipole sum and yields $\Delta\pi=-1$. Note that the leading contribution to the decay width is given by the term with the lowest allowed value of $J$, as the contributions roughly scale as $\sim(kr)^{2J}$. 
Hence, we will only consider $J=2$ in the sum of Eq.~\eqref{eq:GammaEC}.
From parity conservation, see the discussion after Eq.~\eqref{mso4p}, it follows that the axial terms of the weak nuclear current contribute only to $\mathcal{M}_J$, $\mathcal{L}_J$, $\mathcal{T}_J^\text{el}$, while the vector ones contribute only to $\mathcal{T}_J^\text{mag}$.
Within the approximations discussed in App.~\ref{AppA},
one gets from Eqs.~\eqref{lwl1},~\eqref{lwl2} and~\eqref{lwl3}
\begin{gather}
\label{eq:K:el1}
\mathcal{M}_2 \approx 0 \ , \\
\label{eq:K:el2}
\mathcal{L}_2 \approx- \frac{i}{15}  k  F_A  \sum_{j=1}^{A} \hat{T}_{-} \ \vec{\sigma}_j \cdot \vec{\nabla} [r^2Y_{2M}(\hat{r})]_{\vec{r}=\vec{r}_j}   \  , \\
\label{eq:K:el3}
\mathcal{T}_2^{\text{el}} \approx \sqrt{\frac{3}{2}} \, \mathcal{L}_2 \ ,
\end{gather}
while from Eq.~\eqref{lwl4}
\be
\label{eq:K:el4}
\mathcal{T}_2^{\text{mag}} =  \frac{i}{15}\sqrt{\frac{3}{2}}  \frac{k^2}{2m_N} 
\sum_{j=1}^{A} \hat{T}_{-} \    \Big[(F_1+2m_NF_2)\vec{\sigma}_j + \frac{2}{3}F_1 (\vec{r}_j\times\vec{p}_j )\Big]\cdot \vec{\nabla}[r^2Y_{2M}(\hat{r})]_{\vec{r}=\vec{r}_j}  \ ,
\ee
where $k=E_\nu$.
Experimentally~\cite{Wang:2021xhn,KDK:2022hgg}, the neutrino energy and $\ce{^{40}K}$ half-time are respectively $E_\nu\approx43.58(6)$ keV and $T_{1/2}(\ce{^{40}K})=1.266(4) \times10^9$ yr.
Note that the contribution from the vector term is suppressed by a factor $E_\nu/m_N\sim5\times10^{-5}$, 
so it can be safely neglected.
Discarding the experimental uncertainties, which are at the level of per mille, we can extract the value of the nuclear matrix element
\be
\left|\braket{\ce{^{40}Ar}^\star|| \sum_{j=1}^{A} \hat{T}_{-} \ \vec{\sigma}_j \cdot \vec{\nabla} [r^2Y_{2M}(\hat{r})]_{\vec{r}=\vec{r}_j} ||\ce{^{40}K}}\right| \ \approx \ 0.1 \, R_{40} \ ,
\ee
where $R_A\equiv 1.3\text{ fm }A^{1/3}$ is the approximate radius for a nucleus with $A$ nucleons~\cite{book:80102}.

From energy conservation,~\eq{eq:enconsEC}, 
one has $E_\nu = m_i - m_f + \ldots$, 
where the ellipses denote leptonic quantities which, 
to a good approximation, do not depend from $\theta$. 
Hence, the $\theta$-dependence of the neutrino energy 
can be expressed as 
\begin{align}
\label{eq:Enutheta}
E_\nu(\theta)&= E_\nu(0) + [m_i-m_f](\theta) - [m_i-m_f](0) \nonumber \\
&=E_\nu(0)-(\Delta m_N (\theta)-\Delta m_N (0))
=E_\nu(0)+4c_5 \varepsilon M_\pi^2 \left[ \frac{M_\pi^2}{M_\pi^2(\theta)} -1 \right] \ ,
\end{align}
where in the second step the variation of the nuclear BEs from Eq.~\eqref{eq:BEtheta} cancels out in the difference between initial and final nuclei, as it depends only on $A$
which is unchanged in weak decays 
(\emph{cf.}~discussion below Eq.~\eqref{eq:BEtheta}), 
and in the last step we used~\eq{eq:Deltampn}.
Finally, since 
\be
\Gamma_{\text{EC}}\propto E_\nu^4 \ ,
\ee
it follows that
\be
\label{eq:predK}
\frac{\mathring{\Gamma}_{\text{EC}}(0)}{\Gamma_{\text{EC}}(0)}=4\frac{\mathring{E}_{\nu}(0)}{E_{\nu}(0)}\approx18.8 \ .
\ee

 \subsection[$\ce{^{137}Cs}$ $\beta$ decay]{\texorpdfstring{\boldmath{$\ce{^{137}Cs}$ $\beta$}}{137Cs beta} decay}
 
The nucleus of $\ce{^{137}Cs}$, whose half-time is $T_{1/2}(\ce{^{137}Cs})=\qty{30.018(22)}{yr}$, decays to the second excited state of barium $\ce{^{137}Ba^{\star\star}}$ through $\beta^-$-decay the $94.57(26)\%$ of the times~\cite{Tableau}.
The latter then decays to the ground state while emitting a photon with energy $\qty{661.657(3)}{\kilo\electronvolt}$.
The quantum numbers of the initial and final nuclear state are respectively $J_i^{\pi_i}=7/2^-$ and $J_f^{\pi_f}=11/2^+$, while the kinetic energy released in the decay is $Q\equiv m_i-m_f-m_e=\qty{513.97(17)}{\kilo\electronvolt}$~\cite{Wang:2021xhn,Tableau}.
As in the case of $\ce{^{40}K}$ electron capture, one finds the lowest allowed multipole and the parity change to be $J=2$ and $\Delta\pi=-1$,
thus we can rely on the same results of Eqs.~\eqref{eq:K:el1}--\eqref{eq:K:el4} with $k= |\vec{p}_{\nu}+\vec{p}_e|$. The decay width is then obtained by numerically integrating Eq.~\eqref{eq:Gamma:beta}, where again we can safely neglect the contribution from the vector term of the nuclear current, \emph{i.e.} from $\mathcal{T}_2^{\rm mag}$.
Neglecting the small experimental uncertainty, we can extract the value of the nuclear matrix element
\be
\left|\braket{\ce{^{137}Ba^{\star\star}}|| \sum_{j=1}^{A} \hat{T}_{-} \ \vec{\sigma}_j \cdot \vec{\nabla} [r^2Y_{2M}(\hat{r})]_{\vec{r}=\vec{r}_j} ||\ce{^{137}Cs}}\right| \ \approx \ 0.97 \, R_{137} \ .
\ee
Using a similar reasoning as that which led to~\eq{eq:Enutheta}, 
the $\theta$-dependence of the $Q$ value is found to be 
\be
\begin{split}
Q(\theta)= Q(0)+(\Delta m_N (\theta)-\Delta m_N (0)) 
=Q(0)+4c_5 \varepsilon M_\pi^2 \left[ 1 - \frac{M_\pi^2}{M_\pi^2(\theta)} \right] \ ,
\end{split}
\ee
where, again, the variation of the nuclear BEs cancels out in the difference.
Finally, upon numerical integration, we obtain
\label{eq:predCs}
\be
\frac{\mathring{\Gamma}_{\beta^-}(0)}{\Gamma_{\beta^-}(0)}\approx -2.13 \ .
\ee

\section{Underground weak nuclear decay experiments} 
\label{sec:expanddata}

In the last 15 years several $\gamma$-spectroscopy experiments have been made in the underground Gran Sasso Laboratory (LNGS) to search for a possible time dependence of the decay constant of a few nuclei. In this section, we briefly describe two of those experiments, based on 
$\ce{^{40}K}$~\cite{Bellotti:2018jzd} and $\ce{^{137}Cs}$~\cite{Bellotti:2012if}, 
we discuss the obtained results in the 
framework of axion dark matter searches, 
and finally present a new experimental setup currently under 
development. 

We recall here the main advantage of the underground 
environment, which is to suppress the 
varying
background due to cosmic rays. 
Specifically, the
muon and neutron fluxes are respectively suppressed by six and three orders of magnitude, compared to their levels at the Earth's surface, significantly reducing background noise and enhancing the precision of sensitive measurements. 

\subsection[The $\ce{^{40}K}$ experiment]{The \texorpdfstring{\boldmath{$\ce{^{40}K}$}}{40K} experiment}
The $\ce{^{40}K}$ experiment started in September 2015 and ended in November 2017~\cite{Bellotti:2018jzd}. 
A 4 liters 4"$\times$4"$\times$16" NaI crystal detects the \qty{1461}{\kilo\electronvolt} $\gamma$-ray due to the electron capture decay of $\ce{^{40}K}$ to the excited state of $\ce{^{40}Ar}$.
\begin{figure}[t!]
\includegraphics[width=1.0\textwidth,angle=0]{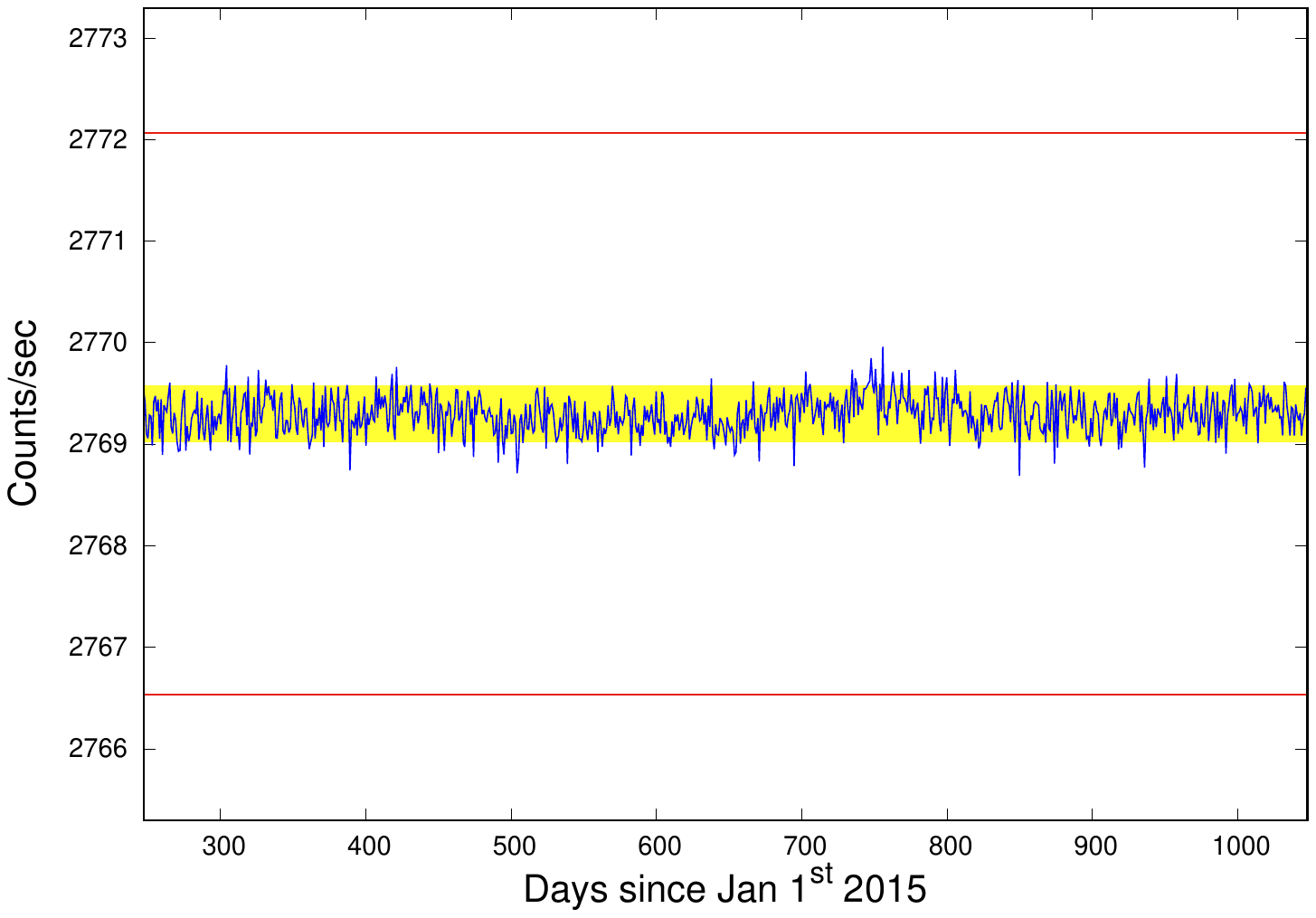}
\vskip -5mm
\caption{The measured rate of the $\ce{^{40}K}$ source averaged over 1 day as function of time. The yellow shaded area corresponds to a $10^{-4}$ uncertainty, the two red lines are at $\pm 10^{-3}$ from the average. 
}
\label{fig:histo40K}
\end{figure}
The source was made by about \qty{9.6}{\kilogram} of potassium bicarbonate powder (KHCO$_{3}$, corresponding to \qty{3.8}{\kilogram} of natural potassium, 
$0.0117\%$ 
of which is $\ce{^{40}K}$), melted with \qty{2.4}{\kilogram} of paraffine grains contained inside a stainless steel box placed around the NaI detector~\cite{Bellotti:2018jzd}, whose light is detected by a photomultiplier. Finally, the signals from the  photomultiplier were processed by an Ortec digiBASE, a 14-pin photomultiplier base connected to the photomultiplier itself.
The whole setup was then shielded by at least 10 cm of lead
and housed in a dedicated
container placed in front
of Hall B of the LNGS.

The intrinsic background, \emph{i.e.}~shielded setup without the KHCO$_{3}$ salt, has been measured during a period of 12 days. Thanks to the underground environment
and to the detector shielding, it was rather low, down to about \qty{6.4}{\hertz} above \qty{15}{\kilo\electronvolt}, to be compared to the source rate
of about \qty{2770}{\hertz}. 
Spectra were stored once per hour.

Fig.~\ref{fig:histo40K} shows the the measured rate of the  source averaged over 1 day 
as a function of time. A time modulation has been searched for by applying the Fourier transform method to the residuals, \emph{i.e.}~the difference between the measured rate and the expected one, for a period of oscillation up to 250 days. For longer periods, the minimization of the chi-squared fit of the residuals was done with a cosine function of time. In short, modulations with amplitude larger than $6.8\times10^{-6}$ ($2\,\sigma$) have been excluded for periods between 6 hours and 800 days.

\subsection[The $\ce{^{137}Cs}$ experiment]{The \texorpdfstring{\boldmath{$\ce{^{137}Cs}$}}{137Cs} experiment}
This $\ce{^{137}Cs}$ experiment took place from June 2011 to January 2012~\cite{Bellotti:2012if}.
The setup was installed in the low background facility STELLA (SubTErranean Low Level Assay) of the LNGS. 
The source consisted in a \qty{3.0}{\kilo\becquerel} $\ce{^{137}Cs}$ standard one with Cs embedded in a plastic disk of 1" diameter and 1/8" thick.
The detector was an Ortec High Purity Germanium with 96$\%$ efficiency, powered by a high voltage power supply connected to the electricity network through an isolation transformer.  
The source was firmly fixed to the copper end-cap of the detector in order to suppress variations in the source-detector relative positions. 
The germanium was surrounded by at least \qty{5}{\centi\meter} of
copper followed by \qty{25}{\centi\meter} of lead to suppress the laboratory gamma ray background. Finally, the shielding and detector were housed in a plastic box, which was flushed with nitrogen at slight overpressure and
worked as an
anti-radon shield. 

The signal from the detector pre-amplifier went first to an amplifier where it was shaped with a \qty{6}{\micro\second} shaping time, and then to a Multi Channel Analyser (Easy-MCA 8k Ortec). 
In order to minimize 
the noise the electronics modules were powered through an insulation transformer. 
The intrinsic background, \emph{i.e.}~shielded detector without the
$\ce{^{137}Cs}$ source, has been measured during a period of 70 days. As expected, it was very low, down to about \qty{40}{counts/\hour} above the \qty{7}{\kilo\electronvolt} threshold, and it added less than \qty{0.01}{\hertz}
to the source rate
of about \qty{700}{\hertz}. The dead time, provided by the MCA, smoothly changed from the initial value of 5.1\% as a consequence of the decreased 
activity of the $\ce{^{137}Cs}$ source.

\begin{figure}[t!]
\centerline{
\includegraphics[width=1.0\textwidth,angle=0]{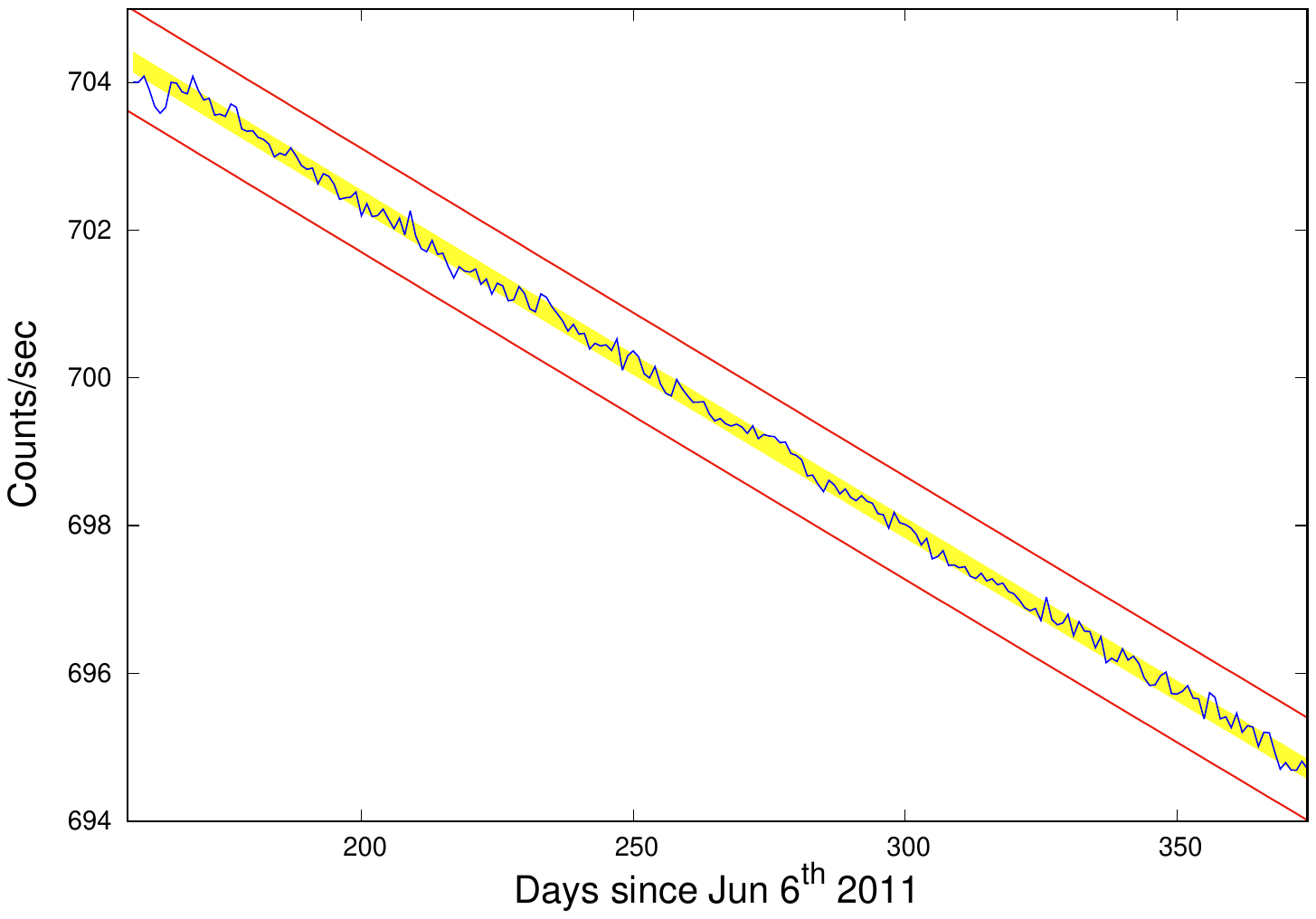}
}
\vskip -5mm
\caption{The measured rate of the $\ce{^{137}Cs}$ source averaged over 1 day as a function of time. The yellow shaded area corresponds to a $2 \times 10^{-4}$ uncertainty,
the two red lines are at $\pm 10^{-3}$ from the average. The first points correspond to the first week of data taking, when
the setup was stabilizing, and they are not considered in the analysis. 
}
\label{fig:histo137Cs}
\end{figure}

Spectra were stored every hour, except during the liquid nitrogen refilling of the detector. These interruptions occurred two times a week and lasted less than 2 hours.
Fig.~\ref{fig:histo137Cs} shows the measured rate of the  source averaged over 1 day as a function of time. A time modulation has been searched for by applying the same method as in the study of $\ce{^{40}K}$ decay. Briefly, modulations with amplitude larger than $9.6\times10^{-5}$ ($2~\sigma$) have been excluded for periods between 6 hours and 1 year.

\begin{figure}[t!]
\centerline{
\includegraphics[width=1.0\textwidth,angle=0]{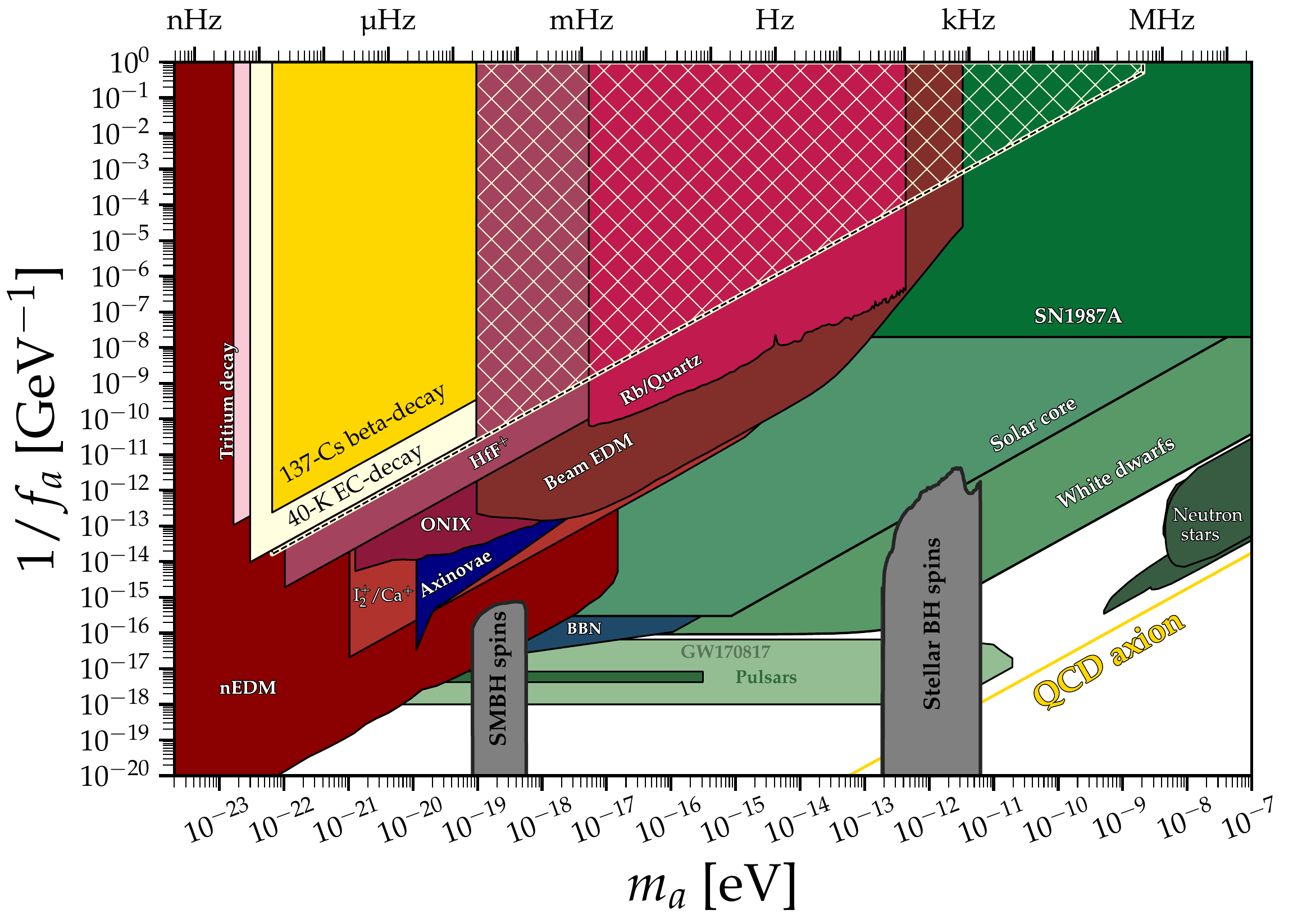}
}
\caption{Constraints on the axion dark matter coupling to gluons. The limits from past measurements are displayed for the $\ce{^{40}K}$ and $\ce{^{137}Cs}$ experiments (yellow-shaded areas), together with the projected sensitivity of the future measurement of $\ce{^{40}K}$ EC decay (yellow cross-hatched area). Limits from laboratory experiments and astrophysics are shown as well for comparison (see text for details). Figure adapted from~\cite{AxionLimits}.}
\label{fig:bounds}
\end{figure}

\subsection{Impact on axion dark matter searches}

The previously obtained limits on the oscillation 
amplitude can be translated on the axion dark matter 
parameter space, employing the theoretical predictions for the 
$\ce{^{40}K}$ and $\ce{^{137}Cs}$ experiments, respectively in~\eq{eq:predK} and~\eq{eq:predCs} -- see also~\eq{eq:Itpred}.
Finding no statistically significant excess in both experiments, we exclude axion decay constants below $1.0\times10^{14}$ 
-- \qty{3.2e10}{\giga\electronvolt} (at $2~\sigma$) for axion masses in the $3.0\times10^{-23}$ -- \qty{9.6e-20}{\electronvolt} range.
The results of our analysis 
are reported in 
\fig{fig:bounds}. 
For comparison, we also display laboratory limits from 
EDM searches~\cite{Abel:2017rtm,Roussy:2020ily,JEDI:2022hxa,Schulthess:2022pbp,Fan:2024pxs}, 
radio-frequency atomic transitions~\cite{Zhang:2022ewz}, 
molecular clocks~\cite{Madge:2024aot} 
and tritium decay~\cite{Zhang:2023lem}, 
as well the model-independent SN 1987A bound~\cite{Springmann:2024ret},\footnote{See also Refs.~\cite{Graham:2013gfa,Lucente:2022vuo} for a weaker 
SN 1987A
bound stemming from the model-independent
axion-nucleon EDM coupling. Note that for 
$1/f_a \gtrsim 3.3 \times 10^{-4} \, \text{GeV}^{-1}$ 
(above the horizontal dashed line in \fig{fig:bounds})
axions enter the trapping regime and the 
cooling bound from SN 1987A does not apply~\cite{Lucente:2022vuo}.}  
finite-density-induced bounds from the solar core and white dwarfs~\cite{Hook:2017psm,Balkin:2022qer}, 
neutron stars cooling~\cite{Gomez-Banon:2024oux,Kumamoto:2024wjd},
gravitational waves~\cite{Hook:2017psm,Zhang:2021mks},
black hole superradiance~\cite{Cardoso:2018tly,Mehta:2020kwu,Baryakhtar:2020gao,Unal:2020jiy,Hoof:2024quk,Witte:2024drg}, 
and other cosmological probes~\cite{Blum:2014vsa,Fox:2023xgx}.

The yellow, QCD axion line stems from the relationship in~\eq{eq:QCDband},   
but it remains beyond the reach of the techniques proposed here.
The standard $m_a$--$f_a$ relation can however be modified
in such a way that the axion mass is suppressed for fixed $f_a$ 
through a symmetry principle~\cite{Hook:2018jle,DiLuzio:2021pxd,DiLuzio:2021gos,Banerjee:2022wzk}.  
This can be achieved by employing 
$\mathcal{N}$ mirror copies of the SM, 
endowed with a $Z_{\mathcal{N}}$ symmetry, 
under which 
$\text{SM}_k \to \text{SM}_{k+1(\text{mod}\, \mathcal{N})}$
and the axion 
acting non-linearly: 
$a \to a + 2\pi k / \mathcal{N}$, with $k = 0,\ldots, \mathcal{N}-1$. 
It can be shown~\cite{Hook:2018jle,DiLuzio:2021pxd} that 
this results in the axion mass being exponentially suppressed as
$z^{\mathcal{N}/2}$, with $z = m_u/m_d \approx 0.5$, 
compared to the usual axion mass.  
Additionally, a modified version of the misalignment mechanism 
can still support the possibility of axion dark matter~\cite{DiLuzio:2021gos}.

\subsection{Future prospects}

We have seen in \sect{sec:timemodwnd} that the $\ce{^{40}K}$ EC decay is more sensitive to an oscillating dark matter background than $\ce{^{137}Cs}$ $\beta$ decay.
Because of this, we decided to prepare a new experiment with a much improved setup for 
$\ce{^{40}K}$ only. In particular, we will have a more efficient shielding by adding \qty{5}{\centi\meter} of OFHC Copper between the source box and the lead (to suppress the $\gamma$-rays due to the  bremsstrahlung from the $\beta$ decay of
the $\ce{^{210}Bi}$ nuclei inside lead) 
and we will change the data acquisition, by introducing 
an event-by-event
acquisition
with \qty{160}{\nano\second} time resolution 
and monitor the time evolution with a Rb clock. In this way, it will be possible to study much shorter timescales, reaching down to \qty{1}{\micro\second}, instead of 6 hours as with the old $\ce{^{40}K}$ setup. As a consequence, we will probe axion masses up to the \qty{1e-9}{\electronvolt} region, with an expected sensitivity ($2~\sigma$) of $4\times10^{-6}$, as displayed in the yellow cross-hatched area in \fig{fig:bounds}. We are planning to start running the new experiment in summer 2025, with the setup housed in the container we have in front of Hall B in the Gran Sasso Laboratory.

There is an additional reason which makes the study of 
$\ce{^{40}K}$ decay at shorter timescales appealing: 
the past observation of non-exponential EC decays of hydrogen-like $\ce{^{140}Pr}$ and $\ce{^{142}Pm}$ at the ion storage ring accelerator facility of GSI Darmstadt. At first, a modulation of about 20$\%$ amplitude and \qty{7}{\second} period of the expected exponential decay curve was observed for both the nuclei~\cite{Litvinov:2008rk}. Subsequently, a new measurement of the $\ce{^{142}Pm}$ lifetime was compatible with a purely exponential decay, with a possible modulation of, at most, about 2$\%$ amplitude~\cite{FRS-ESR:2019pha}. With our new $\ce{^{40}K}$
setup we will be able to test the abovementioned 
parameter space region, in a different EC nuclear decay process, with about four orders of magnitude better sensitivity. 

Regarding alternative choices of nuclei to be employed in the experiment, nuclear processes with lower $Q$-values -- \emph{i.e.}, lower kinetic energy released in the decay 
-- generally exhibit enhanced sensitivity to axion-induced effects. This is because 
the derivative of the logarithmic weak decay rate
approximately scales
inversely with the $Q$-value, following the relation  
$\mathring{\Gamma}/{\Gamma}\propto\mathring{Q}/{Q}\propto\left({\text{MeV}}/{Q}\right)$.
Notable examples of nuclei with very low $Q$-values include $\ce{^{187}Re}$ ($\beta$-decay, $Q=2470.9(13)$ eV~\cite{Filianin:2021nuf}) and $\ce{^{163}Ho}$ (electron capture, $Q=2833\pm30(\text{stat})\pm15(\text{syst})$ eV~\cite{ECHo:2015qgh}).\footnote{Notably, 
$\ce{^{187}Re}$ and $\ce{^{163}Ho}$
were proposed as ideal nuclei for direct neutrino mass searches 
with a sensitivity of $\mathcal{O}(1)$ eV 
\cite{DeRujula:1981ti,DeRujula:1982qt}. Experiments were made with low-temperature micro-calorimeters employing metallic rhenium absorbers \cite{Sisti:2004iq} and a measurement is currently being performed with low-temperature micro-calorimeters implanted with $\ce{^{163}Ho}$ \cite{Borghesi:2023tof}.} These nuclei 
could potentially 
enhance the differential rate, $\mathring{\Gamma}/{\Gamma}$,
by an order of magnitude compared to $\ce{^{40}K}$ and $\ce{^{137}Cs}$.

The ultimate sensitivity of the present experiment is limited by the number of detected events from the radioactive source. Increasing \emph{e.g.}~the source activity by a factor of 10 would enhance the sensitivity on the axion decay constant
by a factor of 3. However, further improvements would necessitate not only a more intense source but also advancements in detection technology. A faster detector, such as 
a lanthanum bromide crystal or
a plastic scintillator, combined with a significantly upgraded data acquisition system, would be essential. Overall, with a setup similar to ours but incorporating 
cutting-edge technologies, an improvement in sensitivity 
on the axion decay constant
of up to two orders of magnitude 
could be achievable.

\section{Conclusions}
\label{sec:concl}

In this work, we have investigated the time modulation of weak nuclear decays as a novel method to explore axion dark matter. By developing a theoretical framework to compute the $\theta$-dependence of weak nuclear decays, including electron capture and $\beta$ decay, we have predicted the time variation of weak radioactivity induced by an oscillating axion dark matter background.
Recasting existing datasets on the weak nuclear decays of $\ce{^{40}K}$~\cite{Bellotti:2018jzd} and $\ce{^{137}Cs}$~\cite{Bellotti:2012if}, collected at the Gran Sasso Laboratory, allowed us to set constraints on the axion decay constant in the mass range of few \qty[print-unity-mantissa = false]{1e-23}{\electronvolt} to \qty[print-unity-mantissa = false]{1e-19}{\electronvolt}. Furthermore, we proposed a new measurement based on $\ce{^{40}K}$ electron capture, enabling the exploration of shorter oscillation periods and axion masses up to \qty[print-unity-mantissa = false]{1e-9}{\electronvolt}.

This study builds upon previous efforts, such as the RadioAxion-$\alpha$ experiment based on $\alpha$-decay of $\ce{^{241}Am}$~\cite{Broggini:2024udi}, and demonstrates the potential of underground laboratories to provide competitive constraints on the axion parameter space. While current sensitivities are moderately weaker than \emph{e.g.}~those of radio-frequency atomic transitions 
and significantly weaker than EDM-based experiments 
(which have the advantage of exhibiting a linear sensitivity to $\theta$), the approach presented in this work offers a better reach compared to existing limits based on radioactivity, such as tritium decay~\cite{Zhang:2023lem}. 

Future efforts will focus on two complementary experimental programs. First, the continued data collection with the $\ce{^{40}K}$ electron capture experiment at the Gran Sasso Laboratory will enable the exploration of a wide range of oscillation periods, from microseconds to few years, providing competitive constraints on the axion parameter space. Second, the ongoing RadioAxion-$\alpha$ experiment~\cite{Broggini:2024udi}, centered on the $\alpha$-decay of $\ce{^{241}Am}$, will complement these efforts by probing similar axion mass ranges with independent nuclear decay channels. Together, these experiments will leverage the unique underground environment to refine our understanding of the $\theta$-dependence in nuclear decays and strengthen the search for axion dark matter.


\section*{Acknowledgments}

We thank Sebastian Hoof for several discussions 
on the topic of this paper, 
as well as 
Matthias Laubenstein and Roberto Menegazzo who participated to the $\ce{^{137}Cs}$ and $\ce{^{40}K}$ experiments.
The work of LDL and CT is supported
by the European Union -- Next Generation EU and
by the Italian Ministry of University and Research (MUR) 
via the PRIN 2022 project n.~2022K4B58X -- AxionOrigins.  JA has received funding from the Fundaci\'on Ram\'on Areces ``Beca para ampliaci\'on de estudios en el
extranjero en el campo de las Ciencias de la Vida y de la
Materia''.

\appendix

\section{Evaluation of the nuclear matrix elements}
\label{AppA}

In this Appendix we provide a few details on the long wavelength approximation and the non-relativistic expansion for nuclear operators employed in the calculation of the nuclear matrix elements of the spherical operators defined in
Eqs.~\eqref{mso1}--\eqref{mso4}.

\subsection{Long wavelength approximation}

The nuclear radius is approximately given by $\sim\qty{1.3}{\femto\meter}\, A^{\frac{1}{3}} = \qty{6.1e-3}{\mega\electronvolt^{-1}} A^{\frac{1}{3}}$~\cite{book:80102}, which implies that in all of the cases of interest the nucleus size is significantly smaller than the wavelength associated with the exchanged momentum $k^{-1}\sim (\qty{1}{\mega\electronvolt})^{-1}$. We can thus expand the spherical Bessel function for small $k r$ as
\begin{equation}
\label{lwl}
j_{J}(kr)\approx\frac{(kr)^{J}}{(2J+1)!!} \ ,
\end{equation}
with higher-order corrections giving a contribution of order $(kr)^2\sim 10^{-3}$ with respect to the leading one for the cases of interest, which can therefore be neglected.
For the spherical operators $\mathcal{M}_{JM}$ and $\mathcal{L}_{JM}$, the expressions of Eqs.~\eqref{mso1} and 
\eqref{mso2} then read
\begin{align}
\label{lwl1}
& \mathcal{M}_{JM} \approx\frac{k^{J}}{(2J+1)!!}\int d^{3}\vec{r}\,r^{J}Y_{JM}\mathcal{J}^{0}(\vec{r}) \, ,\\
\label{lwl12}
& \mathcal{L}_{JM} \approx\frac{1}{i}\frac{k^{J-1}}{(2J+1)!!}\int d^{3}\vec{r}\,r^{J}Y_{JM}\vec{\nabla}\cdot\vec{\mathcal{J}}(\vec{r}) \, .
\end{align}
Note that in the case of a conserved current, {\emph{i.e.}} $\partial^\mu {\cal J}_\mu =0$, a simplification would  occur. Indeed, by assuming the nuclear initial and final state to be eigenstates of the nuclear Hamiltonian, the continuity equation $\vec{\nabla}\cdot\vec{\mathcal{J}}=-\frac{\partial\mathcal{J}^{0}}{\partial t}=-i[H,\mathcal{J}^{0}]$
would yield
\be\label{LtoM}
(m_i - m_f)\braket{N_f||\mathcal{M}_{J}||N_{i}}=k\braket{N_f||\mathcal{L}_{J}||N_{i}} \ ,
\ee
where $m_{i,f}$ are the nuclear masses.
An exception to Eq.~\eqref{lwl12} occurs for the monopole case  $\mathcal{L}_{00}$, since it vanishes at this order. The first contribution thus arises at the next order in the $kr$ expansion and is given by
\begin{equation}
\label{lwl2}
\mathcal{L}_{00}\approx\frac{ik}{6}\int d^{3}\vec{r}\,r^{2}Y_{00}\vec{\nabla}\cdot\vec{\mathcal{J}}(\vec{r}) \ .
\end{equation}
The second order expansion would be also needed for the ${\cal M}_{00}$ monopole expression in the case of a conserved current. This is due to the fact that in this case the integral over space of ${\cal J}^0(\vec r)$ defines the generator $Q$ of the symmetry associated with it. Then, with $| i \rangle$ and $| f \rangle$ being orthogonal eigenstates of the Hamiltonian, one has  $\langle f|Q|i\rangle\propto\langle f|i\rangle=0$. It follows that the first contribution to the operator ${\cal M}_{00}$ is given in this case by
\begin{equation}
\mathcal{M}_{00}\approx -\frac{k^{2}}{6}\int d^{3}\vec{r}\,r^{2}Y_{00}\mathcal{J}^{0}(\vec{r}) \ . 
\end{equation}
Let us now consider the operators $\mathcal{T}_{JM}^{\rm el}$ and $\mathcal{T}_{JM}^{\rm mag}$ of Eqs.~\eqref{mso3} and \eqref{mso4}. By using the identity 
\begin{equation}
\textbf{L}Y_{JM}=-i(\vec{r}\times\vec{\nabla})Y_{JM}=\sqrt{J(J+1)}\textbf{Y}_{JJM} \ ,
\end{equation}
they can be rewritten at the first order in the $kr$ expansion as~\cite{book:80102}
\begin{align}
\label{lwl4}
& \mathcal{T}_{JM}^{\rm mag} \approx \frac{i k^{J}}{(2J+1)!!}\sqrt{\frac{J+1}{J}}\int d^{3}\vec{r}\left\{ \vec{\mu}_{\rm rot}(\vec{r}) + \frac{1}{J+1}\vec{r}\times\vec{J}_{\rm irr}(\vec{r}) \right\}\cdot\vec{\nabla}(r^{J}Y_{JM}) \, , \\
\label{lwl3}
& \mathcal{T}_{JM}^{\rm el}  \approx \frac{1}{i}\frac{k^{J-1}}{(2J+1)!!}\sqrt{\frac{J+1}{J}}\int d^{3}\vec{r}\left\{ \vec{\nabla}\cdot\vec{J}_{\rm irr}(\vec{r}) + \frac{k^{2}}{J+1}\vec{\nabla}\cdot[\vec{r}\times\vec{\mu}_{\rm rot}(\vec{r})] \right\}r^{J}Y_{JM} \, ,
\end{align}
where the vector current has been split into an irrotational field $\vec{\cal J}_{\rm irr}$ and a solenoidal field $\vec{\nabla}\times\vec{\mu}$ as  $\vec {\cal J} = \vec{\cal J}_{\rm irr} + \vec{\nabla}\times\vec{\mu}_{\rm rot}$ in virtue of the Helmholtz's theorem.  For a conserved current,  the matrix element expression of $\mathcal{T}_{J}^{\rm el}$ can be simplified to
\begin{align}
\label{lwl5}
\braket{ N_f|\mathcal{T}_{JM}^{\rm el}|N_{i}}\approx\bra{N_f}\frac{k^{J}}{(2J+1)!!}\sqrt{\frac{J+1}{J}}\int d^{3}\vec{r}&\Biggr\{ \frac{m_i-m_f}{k}r^{J}Y_{JM}\mathcal{J}^{0}(\vec{r}) \nonumber \\
-& \frac{ik}{J+1}\vec{\mu}_{\rm rot}(\vec{r})\cdot[\vec{r}\times\vec{\nabla}(r^{J}Y_{JM})] \Biggr\}\ket{N_{i}} \, ,
\end{align}
again by using the continuity equation. The second term is typically subdominant with respect to the first one, so the latter equation reduces to~\cite{book:80102}
\be
\label{TeltoM}
\braket{ N_f|\mathcal{T}_{JM}^{\rm el}|N_{i}}\approx\sqrt{\frac{J+1}{J}}\frac{m_i-m_f}{k}\braket{ N_f|\mathcal{M}_{JM}|N_{i}} \, .
\ee

\subsection{Non-relativistic expansion for nuclear operators} \label{nonrel}

At the level of quarks, the weak current is given by
\be
\mathcal{J}^{\mu}=\bar{q} \gamma^\mu (1-\gamma_5) \hat{T}_{\pm} q \, ,
\ee
where $\hat{T}_{+}$ ($\hat{T}_{-}$) is the raising (lowering) operator of the approximate $SU(2)$ isospin symmetry acting on the doublet $q =\begin{pmatrix} u & d \end{pmatrix}^T$ of up and down quarks.
The nucleon-nucleon matrix element of the quark current is then
\begin{align}
\braket{p^\prime\sigma^\prime\rho^\prime|\mathcal{J}_{\mu}(0)|p\sigma\rho}=\bar{u}(p^\prime,\sigma^\prime)\Biggr[& F_1 \gamma_\mu + F_2 \sigma_{\mu\nu}k^\nu +iF_S k_\mu \nonumber \\
-&F_A \gamma_\mu\gamma_5 - F_P\gamma_5 k_\mu -F_T \sigma_{\mu\nu}\gamma_5 k^\nu \Biggr] [\hat{T}_{\pm}]_{\rho^\prime\rho} u (p,\sigma) \, ,
\end{align}
where the nucleon states are labeled through their momentum $p^{(\prime)}$, spin $\sigma^{(\prime)}$, and isospin $\rho^{(\prime)}$ quantum numbers.
The conserved vector current hypothesis 
and the CP symmetry enforces $F_S=F_T=0$.
In the $k\to0$ regime, the values of the non-vanishing form factors are $F_1=1$, $2m_N F_2=3.706$ and $F_A=1.27$~\cite{book:80102}, while the Goldberger-Treiman relation yields
\be
F_P=-\frac{2m_N F_A}{k^\mu k_\mu -m^2_\pi}\approx\frac{2m_N F_A}{m^2_\pi} \, .
\ee
Through statistical considerations~\cite{book:14848}, the maximal kinetic energy $E_{\rm c}$ per nucleon in the nucleus is estimated to be around \qty{30}{\mega\electronvolt}, implying that 
a nucleus can be then modeled as a quantum mechanical system of non-relativistic point-like nucleons. One can then take the non-relativistic limit of a nuclear operator and write it in first quantization formalism. The nuclear operator is given by
\be
\mathcal{O}(\vec{r})=\sum_{i=1}^{A}\hat{\mathcal{O}}_{i}^{(1)}(\vec{r}-\vec{r}_{i}) \, ,
\label{eq:nuc_op}
\ee
with the single particle operator $\hat{\mathcal{O}}_{i}^{(1)}(\vec{r}-\vec{r}_{i})\propto\delta(\vec{r}-\vec{r}_{i})$ in the nucleon point-like approximation. We want to match the expression of Eq.~\eqref{eq:nuc_op} 
 with its relativistic counterpart, where the nucleons are described in terms of quantum fields $p(x)$ and $n(x)$ and the nucleon operators are bilinears in $p(x)$ and $n(x)$.
In the case of the weak nuclear current $\mathcal{J}^\mu (\vec{r})$, one gets~\cite{book:80102}
\begin{align}
\mathcal{J}^{0}(\vec{r}) &= \rho(\vec{r}) - \rho_5 (\vec{r})- \frac{1}{i}[H,\phi(\vec{r})] \, , \\
\vec{\mathcal{J}}(\vec{r}) &= \vec{V} (\vec{r}) + \vec{\nabla}\times\vec{\mu}(\vec{r})- \vec{A}(\vec{r}) -\vec{\nabla}\phi(\vec{r}) \, , 
\end{align}
with
\begin{align}
\rho(\vec{r}) &= F_1 \sum_{j=1}^{A}\hat{T}_\pm \delta(\vec{r} - \vec{r}_j) \, , \\
\rho_5(\vec{r}) &= \frac{F_A}{2m_{N}} \sum_{j=1}^{A}\hat{T}_\pm\{\vec{\sigma}_{j}\cdot\vec{p}_{j},\delta(\vec{r} - \vec{r}_j)\} \ , \\
\vec{V}(\vec{r}) &= \frac{F_1}{2m_{N}} \sum_{j=1}^{A}\hat{T}_\pm\{\vec{p}_{j},\delta(\vec{r} - \vec{r}_j)\} \, , \\
\vec{\mu}(\vec{r}) &= \frac{F_1+2m_NF_2}{2m_N} \sum_{j=1}^{A}\hat{T}_\pm\vec{\sigma}_{j}\delta(\vec{r} - \vec{r}_j) \, , \\
\vec{A}(\vec{r}) &= F_A \sum_{j=1}^{A}\hat{T}_\pm\vec{\sigma}_{j}\delta(\vec{r} - \vec{r}_j) \, , \\
\phi(\vec{r}) &= \frac{F_P}{2m_N}  \sum_{j=1}^{A}\hat{T}_\pm \vec{\sigma}_{j}\cdot \vec{\nabla}\left[\delta(\vec{r} - \vec{r}_j)\right] \, . 
\end{align}
The last operator induces contributions that are suppressed by a factor $(k/m_\pi)^2$, so it can be safely neglected.

\begin{small}

\bibliographystyle{utphys}
\bibliography{bibliography.bib}

@article{DeRujula:1981ti,
    author = "De Rujula, A.",
    title = "{A NEW WAY TO MEASURE NEUTRINO MASSES}",
    reportNumber = "CERN-TH-3045",
    doi = "10.1016/0550-3213(81)90002-X",
    journal = "Nucl. Phys. B",
    volume = "188",
    pages = "414--458",
    year = "1981"
}

@article{DeRujula:1982qt,
    author = "De Rujula, A. and Lusignoli, Maurizio",
    title = "{Calorimetric Measurements of $^{163}$Ho Decay as Tools to Determine the Electron Neutrino Mass}",
    reportNumber = "CERN-TH-3377",
    doi = "10.1016/0370-2693(82)90218-0",
    journal = "Phys. Lett. B",
    volume = "118",
    pages = "429",
    year = "1982"
}

@article{Sisti:2004iq,
    author = "Sisti, M. and others",
    title = "{New limits from the Milano neutrino mass experiment with thermal microcalorimeters}",
    doi = "10.1016/j.nima.2003.11.273",
    journal = "Nucl. Instrum. Meth. A",
    volume = "520",
    pages = "125--131",
    year = "2004"
}

@article{Borghesi:2023tof,
    author = "Borghesi, M. and others",
    title = "{An updated overview of the HOLMES status}",
    doi = "10.1016/j.nima.2023.168205",
    journal = "Nucl. Instrum. Meth. A",
    volume = "1051",
    pages = "168205",
    year = "2023"
}

@article{Filianin:2021nuf,
    author = "Filianin, P. and others",
    title = "{Direct Q-Value Determination of the \ensuremath{\beta}- Decay of Re187}",
    eprint = "2108.07039",
    archivePrefix = "arXiv",
    primaryClass = "nucl-ex",
    doi = "10.1103/PhysRevLett.127.072502",
    journal = "Phys. Rev. Lett.",
    volume = "127",
    number = "7",
    pages = "072502",
    year = "2021"
}

@article{ECHo:2015qgh,
    author = "Eliseev, S. and others",
    collaboration = "ECHo",
    title = "{Direct Measurement of the Mass Difference of $^{163}$Ho and $^{163}$Dy Solves the $Q$-Value Puzzle for the Neutrino Mass Determination}",
    eprint = "1604.04210",
    archivePrefix = "arXiv",
    primaryClass = "physics.ins-det",
    doi = "10.1103/PhysRevLett.115.062501",
    journal = "Phys. Rev. Lett.",
    volume = "115",
    number = "6",
    pages = "062501",
    year = "2015"
}

@article{Madge:2024aot,
    author = "Madge, Eric and Perez, Gilad and Meir, Ziv",
    title = "{Prospects of nuclear-coupled-dark-matter detection via correlation spectroscopy of I2+ and Ca+}",
    eprint = "2404.00616",
    archivePrefix = "arXiv",
    primaryClass = "physics.atom-ph",
    doi = "10.1103/PhysRevD.110.015008",
    journal = "Phys. Rev. D",
    volume = "110",
    number = "1",
    pages = "015008",
    year = "2024"
}

@article{Fan:2024pxs,
    author = "Fan, Mingyu and Nima, Bassam and Radak, Aleksandar and Alonso-\'Alvarez, Gonzalo and Vutha, Amar",
    title = "{First results from a search for axionlike dark matter using octupole-deformed nuclei in a crystal}",
    eprint = "2410.02218",
    archivePrefix = "arXiv",
    primaryClass = "physics.atom-ph",
    month = "10",
    year = "2024"
}

@article{Mehta:2020kwu,
    author = "Mehta, Viraf M. and Demirtas, Mehmet and Long, Cody and Marsh, David J. E. and Mcallister, Liam and Stott, Matthew J.",
    title = "{Superradiance Exclusions in the Landscape of Type IIB String Theory}",
    eprint = "2011.08693",
    archivePrefix = "arXiv",
    primaryClass = "hep-th",
    reportNumber = "KCL-PH-TH/2020-77",
    month = "11",
    year = "2020"
}

@article{Gattobigio:2012tk,
    author = "Gattobigio, M. and Kievsky, A. and Viviani, M.",
    title = "{Energy spectra of small bosonic clusters having a large two-body scattering length}",
    eprint = "1206.0854",
    archivePrefix = "arXiv",
    primaryClass = "physics.atm-clus",
    doi = "10.1103/PhysRevA.86.042513",
    journal = "Phys. Rev. A",
    volume = "86",
    pages = "042513",
    year = "2012"
}

@article{Baryakhtar:2020gao,
    author = "Baryakhtar, Masha and Galanis, Marios and Lasenby, Robert and Simon, Olivier",
    title = "{Black hole superradiance of self-interacting scalar fields}",
    eprint = "2011.11646",
    archivePrefix = "arXiv",
    primaryClass = "hep-ph",
    doi = "10.1103/PhysRevD.103.095019",
    journal = "Phys. Rev. D",
    volume = "103",
    number = "9",
    pages = "095019",
    year = "2021"
}

@article{Unal:2020jiy,
    author = {\"Unal, Caner and Pacucci, Fabio and Loeb, Abraham},
    title = "{Properties of ultralight bosons from heavy quasar spins via superradiance}",
    eprint = "2012.12790",
    archivePrefix = "arXiv",
    primaryClass = "hep-ph",
    doi = "10.1088/1475-7516/2021/05/007",
    journal = "JCAP",
    volume = "05",
    pages = "007",
    year = "2021"
}

@article{Hoof:2024quk,
    author = "Hoof, Sebastian and Marsh, David J. E. and Sisk-Reyn\'es, J\'ulia and Matthews, James H. and Reynolds, Christopher",
    title = "{Getting More Out of Black Hole Superradiance: a Statistically Rigorous Approach to Ultralight Boson Constraints}",
    eprint = "2406.10337",
    archivePrefix = "arXiv",
    primaryClass = "hep-ph",
    month = "6",
    year = "2024"
}

@article{Gomez-Banon:2024oux,
    author = "G\'omez-Ba\~n\'on, Antonio and Bartnick, Kai and Springmann, Konstantin and Pons, Jos\'e A.",
    title = "{Constraining Light QCD Axions with Isolated Neutron Star Cooling}",
    eprint = "2408.07740",
    archivePrefix = "arXiv",
    primaryClass = "hep-ph",
    doi = "10.1103/PhysRevLett.133.251002",
    journal = "Phys. Rev. Lett.",
    volume = "133",
    number = "25",
    pages = "251002",
    year = "2024"
}

@article{Cardoso:2018tly,
    author = "Cardoso, Vitor and Dias, \'Oscar J. C. and Hartnett, Gavin S. and Middleton, Matthew and Pani, Paolo and Santos, Jorge E.",
    title = "{Constraining the mass of dark photons and axion-like particles through black-hole superradiance}",
    eprint = "1801.01420",
    archivePrefix = "arXiv",
    primaryClass = "gr-qc",
    doi = "10.1088/1475-7516/2018/03/043",
    journal = "JCAP",
    volume = "03",
    pages = "043",
    year = "2018"
}

@article{Kumamoto:2024wjd,
    author = "Kumamoto, Mia and Huang, Junwu and Drischler, Christian and Baryakhtar, Masha and Reddy, Sanjay",
    title = "{Pi in the Sky: Neutron Stars with Exceptionally Light QCD Axions}",
    eprint = "2410.21590",
    archivePrefix = "arXiv",
    primaryClass = "hep-ph",
    reportNumber = "INT-PUB-24-055",
    month = "10",
    year = "2024"
}

@article{Zhang:2021mks,
    author = "Zhang, Jun and Lyu, Zhenwei and Huang, Junwu and Johnson, Matthew C. and Sagunski, Laura and Sakellariadou, Mairi and Yang, Huan",
    title = "{First Constraints on Nuclear Coupling of Axionlike Particles from the Binary Neutron Star Gravitational Wave Event GW170817}",
    eprint = "2105.13963",
    archivePrefix = "arXiv",
    primaryClass = "hep-ph",
    reportNumber = "Imperial/TP/2021/JZ/01, KCL-PH-TH-2021-25, CERN-TH-2021-061,
  LIGO-P2100161, CERN-TH-2021-061, LIGO-P2100161",
    doi = "10.1103/PhysRevLett.127.161101",
    journal = "Phys. Rev. Lett.",
    volume = "127",
    number = "16",
    pages = "161101",
    year = "2021"
}

@article{Witte:2024drg,
    author = "Witte, Samuel J. and Mummery, Andrew",
    title = "{Stepping Up Superradiance Constraints on Axions}",
    eprint = "2412.03655",
    archivePrefix = "arXiv",
    primaryClass = "hep-ph",
    month = "12",
    year = "2024"
}

@article{Springmann:2024ret,
    author = "Springmann, Konstantin and Stadlbauer, Michael and Stelzl, Stefan and Weiler, Andreas",
    title = "{A Universal Bound on QCD Axions from Supernovae}",
    eprint = "2410.19902",
    archivePrefix = "arXiv",
    primaryClass = "hep-ph",
    reportNumber = "TUM-HEP-1531/24",
    month = "10",
    year = "2024"
}

@article{Litvinov:2008rk,
    author = "Litvinov, Yu. A. and others",
    title = "{Observation of Non-Exponential Orbital Electron Capture Decays of Hydrogen-Like $^{140}$Pr and $^{142}$Pm Ions}",
    eprint = "0801.2079",
    archivePrefix = "arXiv",
    primaryClass = "nucl-ex",
    doi = "10.1016/j.physletb.2008.04.062",
    journal = "Phys. Lett. B",
    volume = "664",
    pages = "162--168",
    year = "2008"
}

@article{FRS-ESR:2019pha,
    author = "Ozturk, F. C. and others",
    collaboration = "FRS-ESR, ILIMA, SPARC, TBWD",
    title = "{New test of modulated electron capture decay of hydrogen-like $^{142}$Pm ions: precision measurement of purely exponential decay}",
    eprint = "1907.06920",
    archivePrefix = "arXiv",
    primaryClass = "nucl-ex",
    doi = "10.1016/j.physletb.2019.134800",
    journal = "Phys. Lett. B",
    volume = "797",
    pages = "134800",
    year = "2019"
}

@article{Broggini:2024udi,
    author = "Broggini, Carlo and Di Carlo, Giuseppe and Di Luzio, Luca and Toni, Claudio",
    title = "{Alpha radioactivity deep-underground as a probe of axion dark matter}",
    eprint = "2404.18993",
    archivePrefix = "arXiv",
    primaryClass = "hep-ph",
    doi = "10.1016/j.physletb.2024.138836",
    journal = "Phys. Lett. B",
    volume = "855",
    pages = "138836",
    year = "2024"
}

@article{Barducci:2022lqd,
    author = "Barducci, Daniele and Toni, Claudio",
    title = "{An updated view on the ATOMKI nuclear anomalies}",
    eprint = "2212.06453",
    archivePrefix = "arXiv",
    primaryClass = "hep-ph",
    doi = "10.1007/JHEP02(2023)154",
    journal = "JHEP",
    volume = "02",
    pages = "154",
    year = "2023",
    note = "[Erratum: JHEP 07, 168 (2023)]"
}

@article{Hook:2017psm,
    author = "Hook, Anson and Huang, Junwu",
    title = "{Probing axions with neutron star inspirals and other stellar processes}",
    eprint = "1708.08464",
    archivePrefix = "arXiv",
    primaryClass = "hep-ph",
    doi = "10.1007/JHEP06(2018)036",
    journal = "JHEP",
    volume = "06",
    pages = "036",
    year = "2018"
}

@article{Balkin:2022qer,
    author = "Balkin, Reuven and Serra, Javi and Springmann, Konstantin and Stelzl, Stefan and Weiler, Andreas",
    title = "{White dwarfs as a probe of light QCD axions}",
    eprint = "2211.02661",
    archivePrefix = "arXiv",
    primaryClass = "hep-ph",
    reportNumber = "IFT-UAM/CSIC-22-136",
    month = "11",
    year = "2022"
}

@article{Stadnik:2013raa,
    author = "Stadnik, Y. V. and Flambaum, V. V.",
    title = "{Axion-induced effects in atoms, molecules, and nuclei: Parity nonconservation, anapole moments, electric dipole moments, and spin-gravity and spin-axion momentum couplings}",
    eprint = "1312.6667",
    archivePrefix = "arXiv",
    primaryClass = "hep-ph",
    doi = "10.1103/PhysRevD.89.043522",
    journal = "Phys. Rev. D",
    volume = "89",
    number = "4",
    pages = "043522",
    year = "2014"
}

@article{Banerjee:2022wzk,
    author = "Banerjee, Abhishek and Eby, Joshua and Perez, Gilad",
    title = "{From axion quality and naturalness problems to a high-quality ZN QCD relaxion}",
    eprint = "2210.05690",
    archivePrefix = "arXiv",
    primaryClass = "hep-ph",
    doi = "10.1103/PhysRevD.107.115011",
    journal = "Phys. Rev. D",
    volume = "107",
    number = "11",
    pages = "115011",
    year = "2023"
}

@article{MadameCurie,
    author = "Curie, M.",
    title = "{PhD Thesis}",
    eprint = "",
    archivePrefix = "",
    primaryClass = "",
    month = "",
    journal = "Doctoral Dissertation, Sorbonne University, Paris",
    year = "1903"
}

@article{McDuffie:2020uuv,
    author = "McDuffie, M. H. and Graham, P. and Eppele, J. L. and Gruenwald, J. T. and Javorsek, D. and Krause, D. E. and Fischbach, E.",
    title = "{Anomalies in Radioactive Decay Rates: A Bibliography of Measurements and Theory}",
    eprint = "2012.00153",
    archivePrefix = "arXiv",
    primaryClass = "nucl-ex",
    month = "11",
    year = "2020"
}

@article{DiLuzio:2020wdo,
    author = "Di Luzio, Luca and Giannotti, Maurizio and Nardi, Enrico and Visinelli, Luca",
    title = "{The landscape of QCD axion models}",
    eprint = "2003.01100",
    archivePrefix = "arXiv",
    primaryClass = "hep-ph",
    reportNumber = "DESY 20-036, DESY-20-036",
    doi = "10.1016/j.physrep.2020.06.002",
    journal = "Phys. Rept.",
    volume = "870",
    pages = "1--117",
    year = "2020"
}

@article{Pomme1,
doi = {10.1088/1681-7575/54/1/1},
url = {https://dx.doi.org/10.1088/1681-7575/54/1/1},
year = {2016},
month = {nov},
publisher = {IOP Publishing},
volume = {54},
number = {1},
pages = {1},
author = {Pomm\'e, S. and others},
title = {On decay constants and orbital distance to the Sun - part I: alpha decay},
journal = {Metrologia}
}

@article{Pomme2,
doi = {10.1088/1681-7575/54/1/19},
url = {https://dx.doi.org/10.1088/1681-7575/54/1/19},
year = {2016},
month = {nov},
publisher = {IOP Publishing},
volume = {54},
number = {1},
pages = {19},
author = {Pomm\'e, S. and others},
title = {On decay constants and orbital distance to the Sun - part II: beta minus decay},
journal = {Metrologia}
}

@article{Pomme3,
doi = {10.1088/1681-7575/54/1/36},
url = {https://dx.doi.org/10.1088/1681-7575/54/1/36},
year = {2016},
month = {nov},
publisher = {IOP Publishing},
volume = {54},
number = {1},
pages = {36},
author = {Pomm\'e, S. and others},
title = {On decay constants and orbital distance to the Sun - part III: beta plus and electron capture decay},
journal = {Metrologia}
}

@misc{AxionLimits,
  author       = {Ciaran O'Hare},
  title        = {cajohare/AxionLimits: AxionLimits},
  month        = jul,
  year         = 2020,
  publisher    = {Zenodo},
  version      = {v1.0},
  doi          = {10.5281/zenodo.3932430},
  howpublished = {\url{https://cajohare.github.io/AxionLimits/}}
}

@article{Abel:2017rtm,
    author = "Abel, C. and others",
    title = "{Search for Axionlike Dark Matter through Nuclear Spin Precession in Electric and Magnetic Fields}",
    eprint = "1708.06367",
    archivePrefix = "arXiv",
    primaryClass = "hep-ph",
    doi = "10.1103/PhysRevX.7.041034",
    journal = "Phys. Rev. X",
    volume = "7",
    number = "4",
    pages = "041034",
    year = "2017"
}

@article{Schulthess:2022pbp,
    author = "Schulthess, Ivo and others",
    title = "{New Limit on Axionlike Dark Matter Using Cold Neutrons}",
    eprint = "2204.01454",
    archivePrefix = "arXiv",
    primaryClass = "hep-ex",
    doi = "10.1103/PhysRevLett.129.191801",
    journal = "Phys. Rev. Lett.",
    volume = "129",
    number = "19",
    pages = "191801",
    year = "2022"
}

@article{Zhang:2022ewz,
    author = "Zhang, Xue and Banerjee, Abhishek and Leyser, Mahapan and Perez, Gilad and Schiller, Stephan and Budker, Dmitry and Antypas, Dionysios",
    title = "{Search for Ultralight Dark Matter with Spectroscopy of Radio-Frequency Atomic Transitions}",
    eprint = "2212.04413",
    archivePrefix = "arXiv",
    primaryClass = "physics.atom-ph",
    doi = "10.1103/PhysRevLett.130.251002",
    journal = "Phys. Rev. Lett.",
    volume = "130",
    number = "25",
    pages = "251002",
    year = "2023"
}

@article{Roussy:2020ily,
    author = "Roussy, Tanya S. and others",
    title = "{Experimental Constraint on Axionlike Particles over Seven Orders of Magnitude in Mass}",
    eprint = "2006.15787",
    archivePrefix = "arXiv",
    primaryClass = "hep-ph",
    doi = "10.1103/PhysRevLett.126.171301",
    journal = "Phys. Rev. Lett.",
    volume = "126",
    number = "17",
    pages = "171301",
    year = "2021"
}

@article{JEDI:2022hxa,
    author = "Karanth, S. and others",
    collaboration = "JEDI",
    title = "{First Search for Axionlike Particles in a Storage Ring Using a Polarized Deuteron Beam}",
    eprint = "2208.07293",
    archivePrefix = "arXiv",
    primaryClass = "hep-ex",
    doi = "10.1103/PhysRevX.13.031004",
    journal = "Phys. Rev. X",
    volume = "13",
    number = "3",
    pages = "031004",
    year = "2023"
}

@article{Fox:2023xgx,
    author = "Fox, Patrick J. and Weiner, Neal and Xiao, Huangyu",
    title = "{Recurrent axion stars collapse with dark radiation emission and their cosmological constraints}",
    eprint = "2302.00685",
    archivePrefix = "arXiv",
    primaryClass = "hep-ph",
    reportNumber = "FERMILAB-PUB-23-029-T",
    doi = "10.1103/PhysRevD.108.095043",
    journal = "Phys. Rev. D",
    volume = "108",
    number = "9",
    pages = "095043",
    year = "2023"
}

@article{Blum:2014vsa,
    author = "Blum, Kfir and D'Agnolo, Raffaele Tito and Lisanti, Mariangela and Safdi, Benjamin R.",
    title = "{Constraining Axion Dark Matter with Big Bang Nucleosynthesis}",
    eprint = "1401.6460",
    archivePrefix = "arXiv",
    primaryClass = "hep-ph",
    doi = "10.1016/j.physletb.2014.07.059",
    journal = "Phys. Lett. B",
    volume = "737",
    pages = "30--33",
    year = "2014"
}

@article{Graham:2013gfa,
    author = "Graham, Peter W. and Rajendran, Surjeet",
    title = "{New Observables for Direct Detection of Axion Dark Matter}",
    eprint = "1306.6088",
    archivePrefix = "arXiv",
    primaryClass = "hep-ph",
    doi = "10.1103/PhysRevD.88.035023",
    journal = "Phys. Rev. D",
    volume = "88",
    pages = "035023",
    year = "2013"
}

@article{Budker:2013hfa,
    author = "Budker, Dmitry and Graham, Peter W. and Ledbetter, Micah and Rajendran, Surjeet and Sushkov, Alex",
    title = "{Proposal for a Cosmic Axion Spin Precession Experiment (CASPEr)}",
    eprint = "1306.6089",
    archivePrefix = "arXiv",
    primaryClass = "hep-ph",
    doi = "10.1103/PhysRevX.4.021030",
    journal = "Phys. Rev. X",
    volume = "4",
    number = "2",
    pages = "021030",
    year = "2014"
}

@article{Sikivie:2020zpn,
    author = "Sikivie, Pierre",
    title = "{Invisible Axion Search Methods}",
    eprint = "2003.02206",
    archivePrefix = "arXiv",
    primaryClass = "hep-ph",
    doi = "10.1103/RevModPhys.93.015004",
    journal = "Rev. Mod. Phys.",
    volume = "93",
    number = "1",
    pages = "015004",
    year = "2021"
}

@article{Irastorza:2018dyq,
    author = "Irastorza, Igor G. and Redondo, Javier",
    title = "{New experimental approaches in the search for axion-like particles}",
    eprint = "1801.08127",
    archivePrefix = "arXiv",
    primaryClass = "hep-ph",
    doi = "10.1016/j.ppnp.2018.05.003",
    journal = "Prog. Part. Nucl. Phys.",
    volume = "102",
    pages = "89--159",
    year = "2018"
}

@article{Weizsacker:1935bkz,
    author = "Weizsacker, C. F. V.",
    title = "{Zur Theorie der Kernmassen}",
    doi = "10.1007/BF01337700",
    journal = "Z. Phys.",
    volume = "96",
    pages = "431--458",
    year = "1935"
}

@article{Donoghue:2006rg,
    author = "Donoghue, John F.",
    title = "{Sigma exchange in the nuclear force and effective field theory}",
    eprint = "nucl-th/0602074",
    archivePrefix = "arXiv",
    doi = "10.1016/j.physletb.2006.10.033",
    journal = "Phys. Lett. B",
    volume = "643",
    pages = "165--170",
    year = "2006"
}

@article{Acharya:2015pya,
    author = "Acharya, Neramballi Ripunjay and Guo, Feng-Kun and Mai, Maxim and Mei\ss{}ner, Ulf-G.",
    title = "{\ensuremath{\theta}-dependence of the lightest meson resonances in QCD}",
    eprint = "1507.08570",
    archivePrefix = "arXiv",
    primaryClass = "hep-ph",
    doi = "10.1103/PhysRevD.92.054023",
    journal = "Phys. Rev. D",
    volume = "92",
    pages = "054023",
    year = "2015"
}

@article{Leutwyler:1992yt,
    author = "Leutwyler, H. and Smilga, Andrei V.",
    title = "{Spectrum of Dirac operator and role of winding number in QCD}",
    reportNumber = "BUTP-92-10",
    doi = "10.1103/PhysRevD.46.5607",
    journal = "Phys. Rev. D",
    volume = "46",
    pages = "5607--5632",
    year = "1992"
}

@article{Brower:2003yx,
    author = "Brower, R. and Chandrasekharan, S. and Negele, John W. and Wiese, U. J.",
    title = "{QCD at fixed topology}",
    eprint = "hep-lat/0302005",
    archivePrefix = "arXiv",
    reportNumber = "DUKE-TH-02-229, MIT-CTP-3201",
    doi = "10.1016/S0370-2693(03)00369-1",
    journal = "Phys. Lett. B",
    volume = "560",
    pages = "64--74",
    year = "2003"
}

@article{Hook:2018jle,
    author = "Hook, Anson",
    title = "{Solving the Hierarchy Problem Discretely}",
    eprint = "1802.10093",
    archivePrefix = "arXiv",
    primaryClass = "hep-ph",
    doi = "10.1103/PhysRevLett.120.261802",
    journal = "Phys. Rev. Lett.",
    volume = "120",
    number = "26",
    pages = "261802",
    year = "2018"
}

@article{DiLuzio:2021gos,
    author = "Di Luzio, Luca and Gavela, Belen and Quilez, Pablo and Ringwald, Andreas",
    title = "{Dark matter from an even lighter QCD axion: trapped misalignment}",
    eprint = "2102.01082",
    archivePrefix = "arXiv",
    primaryClass = "hep-ph",
    reportNumber = "DESY 21-011, DESY-21-011, IFT-UAM/CSIC-20-144, FTUAM-20-21",
    doi = "10.1088/1475-7516/2021/10/001",
    journal = "JCAP",
    volume = "10",
    pages = "001",
    year = "2021"
}

@article{DiLuzio:2021pxd,
    author = "Di Luzio, Luca and Gavela, Belen and Quilez, Pablo and Ringwald, Andreas",
    title = "{An even lighter QCD axion}",
    eprint = "2102.00012",
    archivePrefix = "arXiv",
    primaryClass = "hep-ph",
    reportNumber = "DESY-21-010, DESY 21-010, IFT-UAM/CSIC-20-143, FTUAM-20-21",
    doi = "10.1007/JHEP05(2021)184",
    journal = "JHEP",
    volume = "05",
    pages = "184",
    year = "2021"
}

@article{Dine:1982ah,
	author = "Dine, Michael and Fischler, Willy",
	editor = "Srednicki, M. A.",
	title = "{The Not So Harmless Axion}",
	reportNumber = "UPR-0201T",
	doi = "10.1016/0370-2693(83)90639-1",
	journal = "Phys. Lett. B",
	volume = "120",
	pages = "137--141",
	year = "1983"
}

@article{Abbott:1982af,
	author = "Abbott, L. F. and Sikivie, P.",
	editor = "Srednicki, M. A.",
	title = "{A Cosmological Bound on the Invisible Axion}",
	reportNumber = "PRINT-82-0695 (BRANDEIS)",
	doi = "10.1016/0370-2693(83)90638-X",
	journal = "Phys. Lett. B",
	volume = "120",
	pages = "133--136",
	year = "1983"
}

@article{Preskill:1982cy,
	author = "Preskill, John and Wise, Mark B. and Wilczek, Frank",
	editor = "Srednicki, M. A.",
	title = "{Cosmology of the Invisible Axion}",
	reportNumber = "HUTP-82-A048, NSF-ITP-82-103",
	doi = "10.1016/0370-2693(83)90637-8",
	journal = "Phys. Lett. B",
	volume = "120",
	pages = "127--132",
	year = "1983"
}

@article{Peccei:1977ur,
	author = "Peccei, R. D. and Quinn, Helen R.",
	title = "{Constraints Imposed by CP Conservation in the Presence of Instantons}",
	reportNumber = "ITP-572-STANFORD",
	doi = "10.1103/PhysRevD.16.1791",
	journal = "Phys. Rev. D",
	volume = "16",
	pages = "1791--1797",
	year = "1977"
}

@article{Peccei:1977hh,
	author = "Peccei, R. D. and Quinn, Helen R.",
	title = "{CP Conservation in the Presence of Instantons}",
	reportNumber = "ITP-568-STANFORD",
	doi = "10.1103/PhysRevLett.38.1440",
	journal = "Phys. Rev. Lett.",
	volume = "38",
	pages = "1440--1443",
	year = "1977"
}

@article{Weinberg:1977ma,
	author = "Weinberg, Steven",
	title = "{A New Light Boson?}",
	reportNumber = "HUTP-77/A074",
	doi = "10.1103/PhysRevLett.40.223",
	journal = "Phys. Rev. Lett.",
	volume = "40",
	pages = "223--226",
	year = "1978"
}

@article{Wilczek:1977pj,
	author = "Wilczek, Frank",
	title = "{Problem of Strong  $P$  and  $T$  Invariance in the Presence of Instantons}",
	reportNumber = "Print-77-0939 (COLUMBIA)",
	doi = "10.1103/PhysRevLett.40.279",
	journal = "Phys. Rev. Lett.",
	volume = "40",
	pages = "279--282",
	year = "1978"
}

@article{Abel:2020pzs,
	author = "Abel, C. and others",
	title = "{Measurement of the Permanent Electric Dipole Moment of the Neutron}",
	eprint = "2001.11966",
	archivePrefix = "arXiv",
	primaryClass = "hep-ex",
	doi = "10.1103/PhysRevLett.124.081803",
	journal = "Phys. Rev. Lett.",
	volume = "124",
	number = "8",
	pages = "081803",
	year = "2020"
}

@article{Furnstahl:1999rm,
    author = "Furnstahl, R. J. and Serot, Brian D.",
    title = "{Parameter counting in relativistic mean field models}",
    eprint = "nucl-th/9911019",
    archivePrefix = "arXiv",
    reportNumber = "IU-NTC-99-10",
    doi = "10.1016/S0375-9474(99)00839-8",
    journal = "Nucl. Phys. A",
    volume = "671",
    pages = "447--460",
    year = "2000"
}

@article{Ubaldi:2008nf,
    author = "Ubaldi, Lorenzo",
    title = "{Effects of theta on the deuteron binding energy and the triple-alpha process}",
    eprint = "0811.1599",
    archivePrefix = "arXiv",
    primaryClass = "hep-ph",
    reportNumber = "SCIPP-2008-12",
    doi = "10.1103/PhysRevD.81.025011",
    journal = "Phys. Rev. D",
    volume = "81",
    pages = "025011",
    year = "2010"
}

@article{Damour:2007uv,
    author = "Damour, Thibault and Donoghue, John F.",
    title = "{Constraints on the variability of quark masses from nuclear binding}",
    eprint = "0712.2968",
    archivePrefix = "arXiv",
    primaryClass = "hep-ph",
    doi = "10.1103/PhysRevD.78.014014",
    journal = "Phys. Rev. D",
    volume = "78",
    pages = "014014",
    year = "2008"
}

@article{Bellotti:2012if,
    author = "Bellotti, E. and Broggini, C. and Di Carlo, G. and Laubenstein, M. and Menegazzo, R.",
    title = "{Search for the time dependence of the $^{137}$Cs decay constant}",
    eprint = "1202.3662",
    archivePrefix = "arXiv",
    primaryClass = "nucl-ex",
    doi = "10.1016/j.physletb.2012.02.083",
    journal = "Phys. Lett. B",
    volume = "710",
    pages = "114--117",
    year = "2012"
}

@article{Bellotti:2013bka,
    author = "Bellotti, E. and Broggini, C. and Di Carlo, G. and Laubenstein, M. and Menegazzo, R. and Pietroni, M.",
    title = "{Search for time modulations in the decay rate of $^{40}$K and $^{232}$Th}",
    eprint = "1311.7043",
    archivePrefix = "arXiv",
    primaryClass = "astro-ph.SR",
    doi = "10.1016/j.astropartphys.2014.05.006",
    journal = "Astropart. Phys.",
    volume = "61",
    pages = "82--87",
    year = "2015"
}

@article{Bellotti:2015toa,
    author = "Bellotti, E. and Broggini, C. and Di Carlo, G. and Laubenstein, M. and Menegazzo, R.",
    title = "{Precise measurement of the $^{222}$Rn half-life: A probe to monitor the stability of radioactivity}",
    eprint = "1501.07757",
    archivePrefix = "arXiv",
    primaryClass = "nucl-ex",
    doi = "10.1016/j.physletb.2015.03.021",
    journal = "Phys. Lett. B",
    volume = "743",
    pages = "526--530",
    year = "2015"
}

@article{Bellotti:2018jzd,
    author = "Bellotti, E. and Broggini, C. and Di Carlo, G. and Laubenstein, M. and Menegazzo, R.",
    title = "{Search for time modulations in the decay constant of $^{40}$K and $^{226}$Ra at the underground Gran Sasso Laboratory}",
    eprint = "1802.09373",
    archivePrefix = "arXiv",
    primaryClass = "nucl-ex",
    doi = "10.1016/j.physletb.2018.02.065",
    journal = "Phys. Lett. B",
    volume = "780",
    pages = "61--65",
    year = "2018"
}

@article{Lucente:2022vuo,
    author = "Lucente, Giuseppe and Mastrototaro, Leonardo and Carenza, Pierluca and Di Luzio, Luca and Giannotti, Maurizio and Mirizzi, Alessandro",
    title = "{Axion signatures from supernova explosions through the nucleon electric-dipole portal}",
    eprint = "2203.15812",
    archivePrefix = "arXiv",
    primaryClass = "hep-ph",
    doi = "10.1103/PhysRevD.105.123020",
    journal = "Phys. Rev. D",
    volume = "105",
    number = "12",
    pages = "123020",
    year = "2022"
}

@article{Zhang:2023lem,
    author = "Zhang, Xin and Houston, Nick and Li, Tianjun",
    title = "{Nuclear decay anomalies as a signature of axion dark matter}",
    eprint = "2303.09865",
    archivePrefix = "arXiv",
    primaryClass = "hep-ph",
    doi = "10.1103/PhysRevD.108.L071101",
    journal = "Phys. Rev. D",
    volume = "108",
    number = "7",
    pages = "L071101",
    year = "2023"
}

@article{Lee:2020tmi,
    author = "Lee, Dean and Mei\ss{}ner, Ulf-G. and Olive, Keith A. and Shifman, Mikhail and Vonk, Thomas",
    title = "{\ensuremath{\theta} -dependence of light nuclei and nucleosynthesis}",
    eprint = "2006.12321",
    archivePrefix = "arXiv",
    primaryClass = "hep-ph",
    reportNumber = "FTPI-Minn-20/21, UMN-TH-3922/20",
    doi = "10.1103/PhysRevResearch.2.033392",
    journal = "Phys. Rev. Res.",
    volume = "2",
    number = "3",
    pages = "033392",
    year = "2020"
}

@article{Donoghue:2006du,
    author = "Donoghue, John F.",
    title = "{The Nuclear central force in the chiral limit}",
    eprint = "nucl-th/0603016",
    archivePrefix = "arXiv",
    doi = "10.1103/PhysRevC.74.024002",
    journal = "Phys. Rev. C",
    volume = "74",
    pages = "024002",
    year = "2006"
}

@book{book:17167,
   title =     "Angular momentum in quantum mechanics",
   author =    "A. R. Edmonds",
   publisher = "Princeton University Press"
}

@book{book:80102,
   title =     "Theoretical Nuclear And Subnuclear Physics",
   author =    "John Dirk Walecka",
   publisher = "World Scientific Publishing Company",
   isbn =      "9789812387950,981-238-795-1",
   year =      "2004",
   edition =   "2"
}

@book{book:14848,
   title =     "Quantum mechanics: non-relativistic theory",
   author =    "L. D. Landau and E. M. Lifshitz"
}

@article{Wang:2021xhn,
    author = "Wang, Meng and Huang, W. J. and Kondev, F. G. and Audi, G. and Naimi, S.",
    title = "{The AME 2020 atomic mass evaluation (II). Tables, graphs and references}",
    doi = "10.1088/1674-1137/abddaf",
    journal = "Chin. Phys. C",
    volume = "45",
    number = "3",
    pages = "030003",
    year = "2021"
}

@article{KDK:2022hgg,
    author = "Stukel, M. and others",
    collaboration = "KDK",
    title = "{Rare K40 Decay with Implications for Fundamental Physics and Geochronology}",
    eprint = "2211.10319",
    archivePrefix = "arXiv",
    primaryClass = "nucl-ex",
    doi = "10.1103/PhysRevLett.131.052503",
    journal = "Phys. Rev. Lett.",
    volume = "131",
    number = "5",
    pages = "052503",
    year = "2023"
}

@article{GJORGIEVSKA2024113403,
title = {Revision of the semi-empirical mass formula coefficients by using the ame2020 database},
journal = {Nuclear Engineering and Design},
volume = {426},
pages = {113403},
year = {2024},
issn = {0029-5493},
doi = {https://doi.org/10.1016/j.nucengdes.2024.113403},
url = {https://www.sciencedirect.com/science/article/pii/S002954932400503X},
author = {Sara Gjorgievska and Hristijan Kochankovski and Koviljka Stankovic and Lambe Barandovski}
}

@misc{Tableau,
      title        = {{Tableau du Laboratoire National Henry Becquerell}},
      howpublished = {\url{http://www.lnhb.fr/accueil/donnees-nucleaires/donnees-nucleaires-tableau/}}
}

\end{small}

\end{document}